\documentclass[preprint, superscriptaddress]{revtex4}
\usepackage{graphicx} 
\usepackage{bm} 
\usepackage{multirow} 
\usepackage{subfigure}
\usepackage{amsmath,amsfonts}
\usepackage{color}

\begin{document}
 
\title{Molecular Dynamics Prediction of Thermal Conductivity of GaN Films and Wires at Realistic Length Scales}
 
\author{X. W. Zhou}
\email[]{X. W. Zhou: xzhou@sandia.gov}
\affiliation{Mechanics of Materials Department, Sandia National Laboratories, Livermore, California 94550, USA}

\author{R. E. Jones}
\affiliation{Mechanics of Materials Department, Sandia National Laboratories, Livermore, California 94550, USA}

\author{S. Aubry}
\affiliation{Mechanics and Computation Group, Department of Mechanical Engineering, Stanford University, Stanford, California 94304, USA}

\date{\today}
 
\begin{abstract}

Recent molecular dynamics simulation methods have enabled thermal conductivity of bulk materials to be estimated. In these simulations, periodic boundary conditions are used to extend the system dimensions to the thermodynamic limit. Such a strategy cannot be used for nanostructures with finite dimensions which are typically much larger than it is possible to simulate directly. To bridge the length scales between the simulated and the actual nanostructures, we perform large-scale molecular dynamics calculations of thermal conductivities at different system dimensions to examine a recently developed conductivity vs. dimension scaling theory for both film and wire configurations. We demonstrate that by an appropriate application of the scaling law, reliable interpolations can be used to accurately predict thermal conductivity of films and wires as a function of film thickness or wire radius at realistic length scales from molecular dynamics simulations. We apply this method to predict thermal conductivities for GaN wurtzite nanostructures. 

\end{abstract}
 


\maketitle


\section{Introduction}

Thermal conducting properties of semiconductor nanostructures (e.g., nanowires) has been actively explored in recent years \cite{S2006,BW1998,ZB2001,GJ1999,B2000} because they directly impact many important applications including microelectronics and thermoelectrics. In the microelectronics application, a continuous decrease in feature sizes has resulted in a continuous increase in heat generation. This trend has placed an increasingly demanding requirement for the semiconductor materials to have a high thermal conductivity to effectively dissipate the excessive heat to the surrounding environment \cite{S2006}. In the thermoelectrics \cite{MSS1997} application, on the other hand, a low thermal conductivity is desired because it results in an increase in the energy conversion efficiency. At the nanometer length scale, the effective thermal conductivity becomes sensitive to feature dimensions and defect concentrations. While this provides an effective means to tailor thermal conductivity for specific applications, the scaling of thermal conductivity against feature dimension is not always clear. Because experimental measurement of thermal conductivity is increasingly more challenging as the feature size decreases, a theoretical understanding of thermal conductivity as a function dimension can play a critical role towards optimizing many nanostructure applications including microelectronics and thermoelectrics.

To study the dimension effects on thermal conducting properties of nanostructures, previous works have used the solution of Boltzmann partial differential equations \cite{Zhang2007,LLGC2001,LSC2002,WBKR1999,MB2004,VC1999}. This approach is complex, requiring certain assumptions to reach simple analytical solutions. For example, the simple analytical equations for thermal conductivity provided in references \cite{Zhang2007,LLGC2001,LSC2002} are only applicable either in a small or a large dimension, and hence they cannot be used to extrapolate data obtained from one dimension range to another. In addition, the Boltzmann partial differential equations involve certain input parameters, such as surface specularity, which may not be always available for a given material of interest.

When an accurate interatomic potential is available, the use of molecular dynamics (MD) simulations in studying the thermal transport properties of crystals \cite{KKK2005,CCDG2000,CCG2000,LPY1998,VC2000,LMH1986,VHC1987,MMP1997,OS1999,M1992,PB1994,B1996,SP2001,JJ1999,SPK2002,SPK2004,YCSS2004,WLXO2009,ZAJGS2009} may become desired. This is because the computational system used in MD simulations captures exactly the lattice nature of the crystal, which enables effects of surfaces and defects to be accurately incorporated. It has been shown that a reasonably accurate determination of thermal conductivity requires a real time of MD simulation for at least tens of nanoseconds \cite{ZAJGS2009}. At this time scale, the system size that can be effectively employed usually contains no more than a million of atoms even with massively-parallelized MD simulations. For GaN, this translates to about $1.0 \times10^{4} nm^3$ of material volume. However, the GaN nanowires grown in experiments can have radius exceeding $20 nm$ and length exceeding $20 \mu m$ \cite{SPTS2007}. This corresponds approximately to a material volume exceeding $2.5\times 10^{7} nm^3$. As a result, significant length-scale difference exists between the simulated and the real systems.

Even with rather small systems, MD simulations have been relatively successfully applied to determine thermal conductivities of bulk materials based upon either the Green-Kubo (and its variations) \cite{KKK2005,CCDG2000,CCG2000,LPY1998,VC2000,LMH1986,VHC1987} or the ``direct method'' \cite{MMP1997,OS1999,M1992,PB1994,B1996,SP2001,JJ1999,SPK2002,SPK2004,YCSS2004,WLXO2009,ZAJGS2009}. In the Green-Kubo method, periodic boundary conditions are used in all the three coordinate directions. As was demonstrated in previous work \cite{ZAJGS2009} and will be reexamined in the following, the use of periodic boundary conditions effectively extends the dimension of a small computational system to infinity. As a result, an infinitely large bulk crystal can be well captured with the Green-Kubo method even when a small simulated system is used. In the ``direct method'', periodic boundary conditions can also be applied in the two coordinate directions perpendicular to the heat flux so that these two directions can be viewed as infinity. However, the ``direct method'' involves a heat source and heat sink along the heat conducting direction. A finite spacing, $L$, must be imposed between the source and the sink. Fortunately, both experiments and theories indicated that the inverse of thermal conductivity $1/\kappa$ and the inverse of length scale $1/L$ satisfy accurately a linear scaling relationship \cite{LHM2008,H2007,SPK2002,M1992,OS1999,PB1994}:
\begin{equation} \frac{1}{\kappa} = \frac{1}{\kappa_b}+\frac{\alpha}{L}
\label{previous extrapolation}
\end{equation}
where $\kappa_b$ is the bulk thermal conductivity at $L \rightarrow \infty$ , and $\alpha$ is a dimension-independent coefficient. To obtain bulk thermal conductivity, several simulations for different small cell lengths are performed, and the results can then be relatively accurately extrapolated to the infinite-size limit due to the linearity of the relationship. 

Periodic boundary conditions cannot be used for finite system dimensions. However, if a reliable linear scaling law that is applicable from nano up to macro scales is known, the thermal conductivity of finite systems at realistic length scales can still be accurately predicted based upon data obtained from MD simulations on a nano scale. An underlying assumption of Eq. (\ref{previous extrapolation}) is that the two dimensions perpendicular to the heat flux are infinite. Because of this, Eq. (\ref{previous extrapolation}) is essentially a scaling law for 2D films where the heat flux is through the film thickness $L$. Unfortunately, no similar scaling laws were previously available for other heat flux directions (e.g., parallel to the film surface) or other nanostructures (e.g., nanowire or nanoparticles). As a result, previous MD simulations had not been applied to calculate thermal conductivities at realistic length scales for many interesting nanostructures, including cases where heat conduction occurs in the plane of a film or through the axis of a wire \cite{WLXO2009,PSM2007}. 
 
Recently, we developed a theoretical scaling law that defines thermal conductivity of a nanostructure as a function of all of its three independent dimensions: thickness $t$, width $W$, and length $L$ \cite{ZJA2009}:
\begin{eqnarray}
&\kappa&\left(t,W,L\right) = \frac{L \cdot \kappa_{0,c}}{L+\delta_0} - \left[\frac{L \cdot \kappa_{0,c}}{L+\delta_0} - \frac{L \cdot \kappa_{1,c}}{L+\delta_1}\right] \cdot \left(\frac{2d}{t} + \frac{2d}{W}\right) \nonumber \\ &+& \left[\frac{L \cdot \kappa_{0,c}}{L+\delta_0} + \frac{L \cdot \kappa_{2,c}}{L+\delta_2} - 2 \cdot \frac{L \cdot \kappa_{1,c}}{L+\delta_1}\right] \cdot \frac{4d^2}{t \cdot W}
\label{general law}
\end{eqnarray}
where $\kappa_{0,c}$, $\kappa_{1,c}$, $\kappa_{2,c}$, $\delta_0$, $\delta_1$, $\delta_2$, and $d$ are seven constants that can in principle be determined from available thermal conductivity vs. dimension data. By performing very large MD simulations at different system dimensions, we have demonstrated that Eq. (\ref{general law}) is highly accurate from a nano scale all the way to macro scales \cite{ZJA2009}. The development of such an analytical scaling law has begun to enable MD simulations to be used to predict thermal conductivity of nanostructures at realistic length scales. The goal of the present work is threefold. First, we provide more detailed physics of the scaling law by adapting it for general 2D film and 1D wire cases. Next, we explore the conditions and parameter space under which the scaling law can be accurately applied, and discuss the methods to predict thermal conductivity of nanostructures at realistic length scales. Finally, we perform large scale MD simulations to determine the $[0001]$ thermal conductivities of a wurtzite GaN crystal constructed in two nanostructure configurations: (i) $(1\bar100)$ film with varying film thickness and (ii) a $[0001]$ hexagonal wire with varying wire radius. GaN is chosen for the case study because it has excellent optoelectronic properties and can be easily integrated with the existing silicon structures. In addition, some GaN applications, such as laser diodes and high electron mobility transistors \cite{JCKSYS2002,ZQWL2003,KCLKRKKC2004,QLGWBL2004,HDCL2002,CJHLKPGSY2003}, operate at high current and power densities. Understanding thermal transport of GaN nanostructures helps these applications.

\section{Scaling law}

The underlying assumption of our scaling law \cite{ZJA2009} is accurate when the dimension of the structure is larger than the phonon mean free path. For GaN bulk crystals, the phonon mean free path has been estimated to be approximately 500 \AA ~at 300 K and 100 \AA ~at 500 K from both experimental data and kinetic theory \cite{DPWJP2006}. In nanostructures, the apparent mean free path is reduced due to surface scattering, causing the thermal conductivity to reduce. Interestingly, for given cross section, simple theoretical analysis indicated that the inverse of the phonon mean free path along the length direction is a linear function of inverse of the length \cite{SPK2002}. This suggests that the inverse of the thermal conductivity is also a linear function of the inverse of the length, matching exactly the prediction of the scaling law. This means that the scaling law can actually be applied even when the length scale is comparable with the phonon mean free path. We will reexamine this in the following.

Our theory can be extended and applied to arbitrary heat flux directions with respect to arbitrary shapes of the nanostructure. Here we confine our discussion to film and wire cases.

\subsection{The film case}

\begin{figure}
\includegraphics[width=6in]{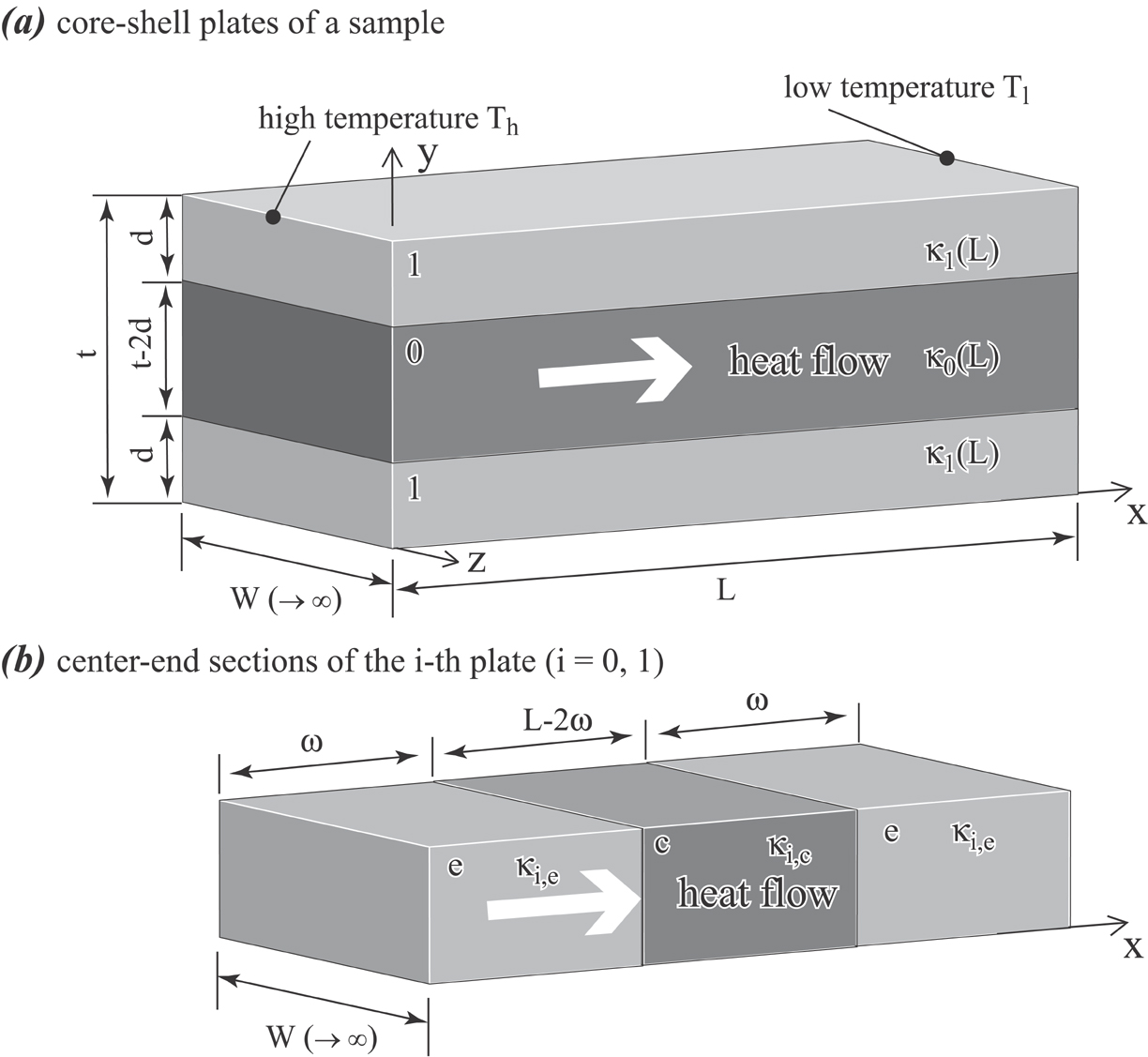}
\caption{Heat conduction through a finite length $L$ of a box-shaped film with a finite thickness $t$ and an infinite width $W \rightarrow \infty$.
\label{film model}}
\end{figure}

The geometry of the film case is illustrated in Fig. \ref{film model}(a) where heat is assumed to flow through a finite length $L$ of a box-shaped sample with a finite thickness $t$ and an infinite width $W \rightarrow \infty$. Note that for a true 2D film, $L \rightarrow \infty$. A more general scenario of finite $L$ is assumed here so that the theory can be applied with the direct method MD simulations where a finite spacing between heat source and heat sink must be used. It is recognized that the size-effect on thermal conductivity origins from the surface scattering of phonons. Hence, we separately consider surface and bulk regions of the sample. As shown in Fig. \ref{film model}(a), the sample is divided along the thickness direction into three smaller box-shaped regions (referred to as plates hereafter): the inner (core) plate has a thickness of $t - 2d$ and is marked as ``0'' because it does not bound any y- surfaces, and the two outer (shell) plates have a thickness $d$ and are designated as ``1'' because they bound one y- surface. When the thickness $t$ is very large, we can always choose a sufficiently large shell thickness $d$ so that the thermal transport behavior of plate 0 is independent of the presence of the top and the bottom free surfaces that are far away. This means that the thermal conductivity of plate 0 is independent of $t$ and therefore can be expressed as a function of $L$ only: $\kappa_0\left(L\right)$. The local thermal conductivity inside plate 1 is nonuniform near the surface. However, plate 1 still exhibits an apparent overall thermal conductivity. Note that at a large $d$, there is really no ``distinguishable'' interface between plates 0 and 1 as the thermal transport properties from both sides of the interface approach the same bulk values. This means that once a large value of $d$ is given, the apparent thermal conductivity of plate 1 can also be expressed as a function of $L$ only: $\kappa_1\left(L\right)$.

Using Fig. \ref{film model}(a), we assume that the left hand side of the sample is kept at a high temperature of $T_h$ and the right hand side at a low temperature of $T_l$. Because the vertical temperature gradient is zero at the ``indistinguishable'' interface between plates 0 and 1, we can list separately the thermal transport equations for the two types of plates:
\begin{equation}
\kappa_0\left(L\right) = \frac{J_0}{\frac{T_h - T_l}{L}}
\label{kappa_0_1}
\end{equation}
\begin{equation}
\kappa_1\left(L\right) = \frac{J_1}{\frac{T_h - T_l}{L}}
\label{kappa_0_2}
\end{equation}
where $J_0$ and $J_1$ are, respectively, the heat fluxes through plates 0 and plate 1. Note that because of an assumed zero vertical temperature gradient at the 0/1 interface, the high and low temperatures are the same for both types of plates. The overall thermal conductivity of the system is expressed as
\begin{equation}
\kappa = \frac{J}{\frac{T_h - T_l}{L}}
\label{kappa 1}
\end{equation}
where the total flux $J$ can be calculated as an area-weighted average
\begin{equation}
J = \frac{W \cdot \left(t - 2d \right) \cdot J_0 + 2W \cdot d \cdot J_1}{W \cdot t} = \frac{\left(t - 2d \right) \cdot J_0 + 2d \cdot J_1}{t} 
\label{J 1}
\end{equation}
as in parallel conductors.
Substituting Eqs. (\ref{kappa_0_1}), (\ref{kappa_0_2}), and (\ref{J 1}) into Eq. (\ref{kappa 1}), we have
\begin{equation}
\kappa\left(t,L\right) = \kappa_0\left(L\right) - \left[\kappa_0\left(L\right) - \kappa_1\left(L\right)\right] \cdot \frac{2d}{t}
\label{kappa 2}
\end{equation}

Now we consider the thermal transport through the $i$-th plate ($i = 0, 1$). Imagine that the plate is divided along the length direction into three sections: the center section contains a length of $L - 2\omega$ and is marked as ``c'', and the two end sections contain a length of $\omega$ and are marked as ``e'', as shown in Fig. \ref{film model}(b). Just as a subsurface thickness $d$ subsumes the scattering of the side surfaces, a subsurface length $\omega$ subsumes the scattering of the end surfaces (including the artificial effects of the thermostats). Here we distinguish $d$ and $\omega$ for generality. It can be seen that for a given large value of $\omega$, which is always possible when $L$ is sufficiently large, the thermal transport behavior of the center section is independent of the presence of the two end surfaces that are far away. This means that the apparent thermal conductivity of the center section is equal to a constant $\kappa_{i,c}$ ($i = 0, 1$). In particular, $\kappa_{0,c}$ corresponds to the bulk thermal conductivity $\kappa_b$ by definition. Similar to the discussion in the above, the apparent thermal conductivity exhibited by the two end sections is also independent of $L$ and therefore is equal to another constant $\kappa_{i,e}$. Because heat flows through the three sections of the plate in serial, the heat flux J is a constant. We can list the temperature difference between the left and right ends as:
\begin{equation}
T_h - T_l = \frac{J}{\kappa_{i,c}} \cdot \left(L - 2\omega\right) + \frac{J}{\kappa_{i,e}} \cdot \left(2\omega\right)
\label{delta T 1}
\end{equation}
The inverse of the overall thermal conductivity of the plate, $\kappa_i^{-1}\left(L\right)$, equals $J^{-1} \cdot \left(T_h - T_l\right)/L$. We can therefore write: 
\begin{equation}
\frac{1}{\kappa_i\left(L\right)} = \frac{1}{\kappa_{i,c}} + \frac{2\omega}{L} \cdot \left( \frac{1}{\kappa_{i,e}} - \frac{1}{\kappa_{i,c}} \right)
\label{resistivity 1}
\end{equation}
Eq. (\ref{resistivity 1}) can be rewritten as
\begin{equation}
\kappa_i\left(L\right) = \frac{L \cdot \kappa_{i,c}}{L+\delta_i}
\label{kappa i 1}
\end{equation}
where
\begin{equation}
\delta_i = 2\omega \cdot \frac{\kappa_{i,c}-\kappa_{i,e}}{\kappa_{i,e}}
\label{delta i 1}
\end{equation}
$\delta_i$ combines the relative change of thermal conductivities between the center and the end sections with the length $\omega$, it therefore reduces one parameter. This reduction in parameters is expected because $\omega$ and $\kappa_{i,e}$ are dependent. It can be seen from Eq. (\ref{delta i 1}) that $\delta_i$ can be viewed as a characteristic length measuring the scattering of the end surfaces. Substituting Eq. (\ref{kappa i 1}) into Eq. (\ref{kappa 2}), we have a scaling law for the thin film:
\begin{equation}
\kappa\left(t,L\right) = \frac{L \cdot \kappa_{0,c}}{L+\delta_0} - \left[\frac{L \cdot \kappa_{0,c}}{L+\delta_0} - \frac{L \cdot \kappa_{1,c}}{L+\delta_1}\right] \cdot \frac{2d}{t}
\label{kappa 3}
\end{equation}

Eq. (\ref{kappa 3}) is consistent with the previous work \cite{ZAJGS2009}. For instance, it reduces to Eq. (\ref{previous extrapolation}) when $t \rightarrow \infty$, and it matches Eq. (\ref{general law}) when $W \rightarrow \infty$. Eq. (\ref{kappa 3}) involves five parameters $\kappa_{0,c}$, $\kappa_{1,c}$, $\delta_0$, $\delta_1$, and $d$. These five parameters have physical meanings and must be subject to some physical constraints. First, the surface thickness $d$ and the associated surface thermal conductivity $\kappa_{1,c}$ are dependent parameters. Hence, $d$ can be selected and only the corresponding $\kappa_{1,c}$ value be treated as an unknown parameter. However, $d$ is not completely arbitrary as it must be large enough to subsume the surface scattering effect. Once $d$ is large enough, Eq. (\ref{kappa 3}) can always predict accurate results regardless of its particular value for any film thickness $t$ that satisfies the geometry condition $t > 2d$. On the other hand, a large $d$ prevents Eq. (\ref{kappa 3}) from being used for small thickness $t$ due to the constraint $t > 2d$. So it is important to understand the low bound of $d$. Clearly $d$ is sufficiently big if it equals the phonon mean free path in the bulk crystal. As described above, this might be overly stringent.

Once $d$ is chosen, the remaining parameters can be fitted to the available data. For MD applications, it is necessary to perform several simulations at different dimensions in order to fit Eq. (\ref{kappa 3}). Note that if the minimum system thickness used in these simulations is $t_{min}$, then the largest $d$ that still enables all the MD data to satisfy the geometry condition is $t_{min}/2$. In order to find a small $d$ to enable study of small structures, a trial-and-error approach can be used. For instance, $d$ can be first set to $t_{min}/2$, and Eq. (\ref{kappa 3}) fitted to all MD data. If satisfactory fitting is obtained as will be described below, then the selected $d$ is good. Otherwise $d$ can be set according to the next thinnest sample, and the thinnest ($t_{min}$) sample is disqualified from the fitting. This process is continued until an appropriate $d$ is found. There are also some useful relations. Because the end section is assumed to have more surface scattering than the center section, and plate 1 has surface scattering that is assumed to be insignificant in plate 0, we always have $\delta_i > 0$ (i = 0, 1), $\kappa_{1,c} < \kappa_{0,c}$, and $\kappa_1\left(L\right) < \kappa_0\left(L\right)$ (for any $L$). These conditions are automatically satisfied during fitting provided that the data to be fit are accurate and $d$ satisfies the geometry constraint. 

Eq. (\ref{kappa 3}) can be used for infinite 2D films. When $t \rightarrow \infty$, the problem is essentially the heat conduction through the length $L$ of a film ($L$ is in fact the ``thickness'' in this case). When $L \rightarrow \infty$, Eq. (\ref{kappa 3}) gives thermal conductivity in the plane of a film as a function of the film thickness $t$. In addition, Eq. (\ref{kappa 3}) can also be used for quasi- 2D cases or even 1D cases, e.g. out-of-plane conduction, to explore the dimensional effects by using different $t/L$ ratios. 
 
\subsection{The wire case}

\begin{figure}
\includegraphics[width=6in]{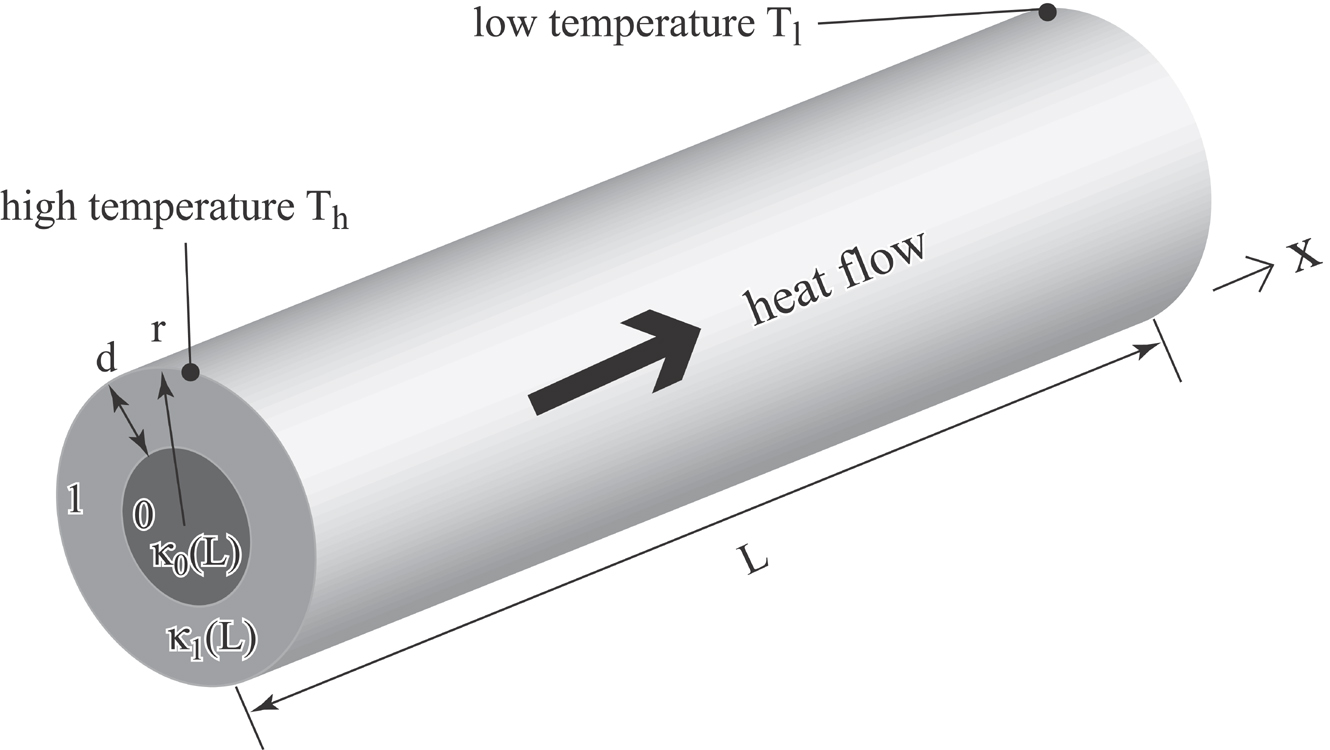}
\caption{Heat conduction through a finite length $L$ of a circular wire with a finite radius $r$.
\label{wire model}}
\end{figure}

The wire case is illustrated in Fig. \ref{wire model} where heat is assumed to flow through the length $L$ of a circular sample with a finite radius $r$. Using the same theory described above, the sample is divided along radius direction into an inner, smaller cylindrical core with a radius of $r - d$ and an outer cylindrical shell with a thickness of $d$. The core does not bound any free surfaces whereas the shell terminates with a radial surface, and hence the core and shell are denoted as ``0'' and ``1'' respectively. It can be seen that when $r$ is very large, we can always choose a sufficiently large $d$ to subsume the surface scattering effect so that the thermal conductivities of the core and the shell are independent of the wire radius and hence can be expressed as functions of $L$ using $\kappa_0\left(L\right)$ and $\kappa_1\left(L\right)$ respectively. In Fig. \ref{wire model}, we again assume that the sample is held at a high temperature of $T_h$ at the left and a low temperature of $T_l$ at the right. The thermal transport equations for the core, shell and overall system can be represented by Eqs. (\ref{kappa_0_1}) - (\ref{kappa 1}). The total flux $J$, however, is modified as
\begin{equation}
J = \frac{J_0 \cdot \pi \cdot \left(r - d \right)^2 + J_1 \cdot \left[\pi \cdot r^2 - \pi \cdot \left( r - d \right)^2 \right]}{\pi \cdot r^2} = \frac{J_0 \cdot \left(r - d \right)^2 + J_1 \cdot d \left( 2 r - d \right)}{r^2} 
\label{J 2}
\end{equation}
Substituting Eqs. (\ref{kappa_0_1}), (\ref{kappa_0_2}), and (\ref{J 2}) into Eq. (\ref{kappa 1}), we have
\begin{equation}
\kappa\left(r,L\right) = \kappa_0\left(L\right) - \left[\kappa_0\left(L\right) - \kappa_1\left(L\right)\right] \cdot \left(\frac{2d}{r} - \frac{d^2}{r^2}\right)
\label{kappa length radius 2}
\end{equation}
where $\kappa_0\left(L\right)$ and $\kappa_1\left(L\right)$ can be described by Eq. (\ref{kappa i 1}). Substituting Eq. (\ref{kappa i 1}) into Eq. (\ref{kappa length radius 2}), we have a scaling law for the wire:
\begin{equation}
\kappa\left(r,L\right) = \frac{L \cdot \kappa_{0,c}}{L+\delta_0} - \left(\frac{L \cdot \kappa_{0,c}}{L+\delta_0} - \frac{L \cdot \kappa_{1,c}}{L+\delta_1} \right) \cdot \left(\frac{2d}{r} - \frac{d^2}{r^2}\right)
\label{kappa 4}
\end{equation}

Eq. (\ref{kappa 4}) is also consistent with the previous work \cite{ZAJGS2009} as it reduces to Eq. (\ref{previous extrapolation}) when $r \rightarrow \infty$, and it matches Eq. (\ref{general law}) using the geometry conditions of a circular wire \cite{ZJA2009}: $t = W = 2r$, $\kappa_{2,c} = \kappa_{1,c}$, and $\delta_2 = \delta_1$. Eq. (\ref{kappa 4}) involves the same five parameters as in the film case. The geometry of the wire case, however, requires that $d < r$. For MD applications, the maximum $d$ enabling all MD data equals the minimum radius $r_{min}$ used in the series of MD simulations. With $d$ determined similarly as in the film case, the remaining four parameters can be fitted to the measurements. If done correctly, the parameters should satisfy $\delta_i > 0$ (i = 0, 1), $\kappa_{1,c} < \kappa_{0,c}$, and $\kappa_1\left(L\right) < \kappa_0\left(L\right)$ (for any $L$). 

Eq. (\ref{kappa 4}) can have numerous uses. When $L \rightarrow \infty$, Eq. (\ref{kappa 4}) represents thermal conductivity through an infinite 1D wire as a function of wire radius. In particular, Eq. (\ref{kappa 4}) indicates that thermal conductivity of wires is a linear function of $2d/r - d^2/r^2$. When $r$ is large, the thermal conductivity increases to a first order with $\sim -1/r$, in agreement with the approximate equation derived by Lu et al from the Boltzmann equation \cite{LLGC2001,LSC2002}. Eq. (\ref{kappa 4}) can also be used in other cases. For instance, at $r \rightarrow \infty$, the problem reduces to heat conduction through the thickness $L$ of an infinite 2D film. It can be used for quasi- 2D films or even 3D particles to explore the dimension effects by using different $r/L$ ratios.

\section{Molecular dynamics methods}

One ultimate goal of our work is to enable MD simulations to predict thermal conductivities of GaN films and wires at realistic, device length scales on the order of $100-1000$ \AA ~or more, which is not at bulk limit but too long to directly simulate with MD. Here we describe details of the interatomic potential used in the MD, the computational cells for film and wire configurations, and the thermal transport simulation method.
 
\subsection{Interatomic potential}

The previous work \cite{ZAJGS2009,ZJA2009} applied the Stillinger-Weber (SW) potential parameterized by Bere and Serra \cite{BS2002,BS2006} to calculate the thermal conductivity of GaN bulk crystals. To compare with the previous results, we also use the same potential in the present study. This potential gives reasonable prediction on dispersion relations, vibrational density of states (DOS), and heat capacity for bulk systems \cite{ZAJGS2009}. 

\subsection{Computational system for films}

\begin{figure}
\includegraphics[width=4in]{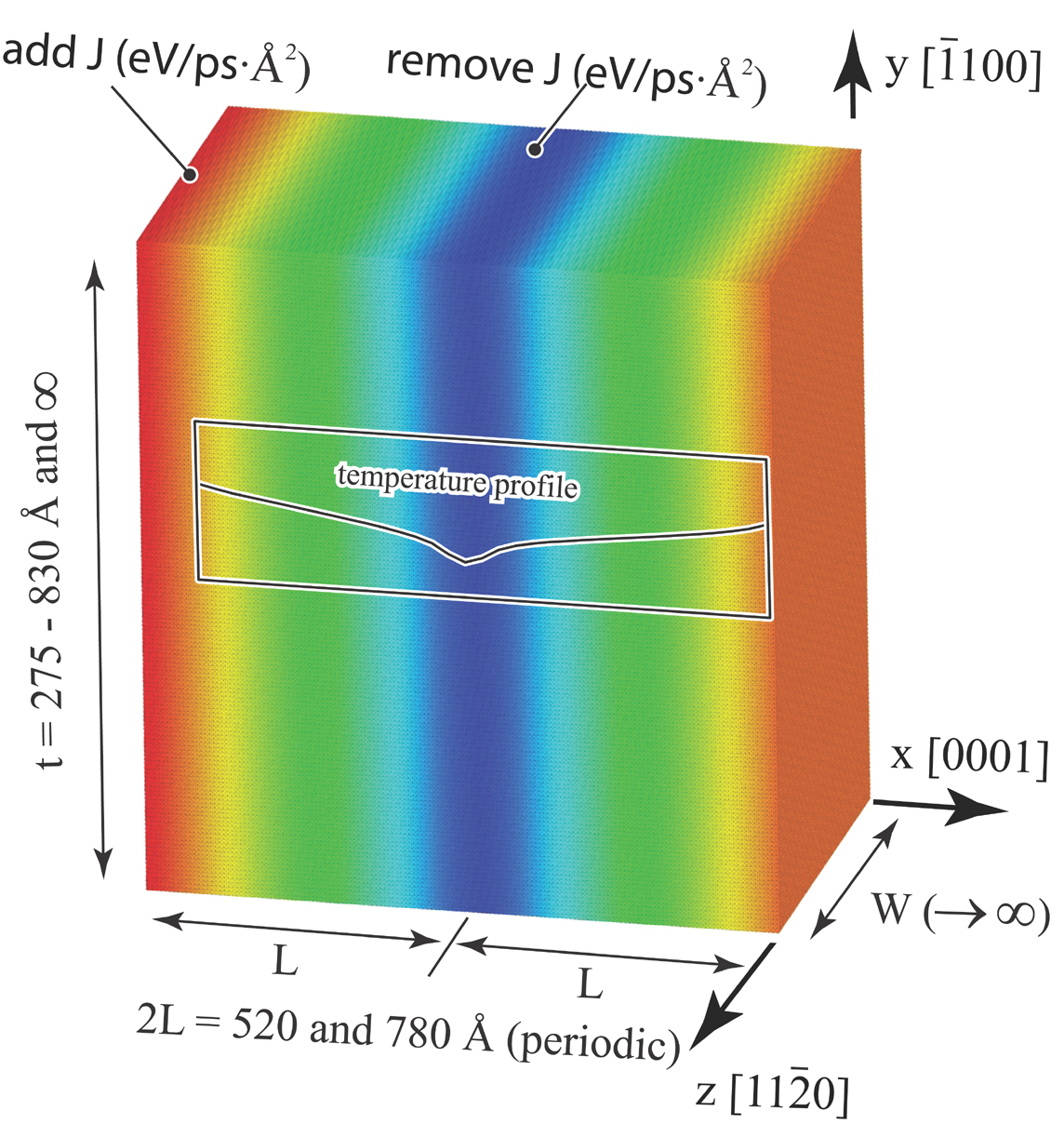}
\caption{Atomistic configuration for film MD simulations. 
\label{thin film MD model}}
\end{figure}

The computational system used for the film simulations is shown in Fig. \ref{thin film MD model}, where the color scheme shows the temperature (red means the highest temperature and blue means the lowest temperature). Similar to Fig. \ref{film model}, we assume that the sample has a finite thickness $t$ in the y- direction and an infinite width $W \rightarrow \infty$ in the z- direction. Following the customary approach with the direct method MD simulations \cite{MMP1997,OS1999,M1992,PB1994,B1996,SP2001,JJ1999,SPK2002,SPK2004,YCSS2004,WLXO2009,ZAJGS2009}, a periodic boundary condition was used in the x- direction. As will be shown below, this means that the system dimension in the x- direction is $2L$ rather than $L$.   

The equilibrium GaN has a wurtzite hexagonal crystal structure. The hexagonal crystal has three orthogonal directions $[0001]$, $[\bar1100]$, and $[11\bar20]$. To study thermal conduction along the $[0001]$ direction of a low energy $(\bar1100)$ film, the computational supercell is aligned so that the $x-$, $y-$, and $z-$ coordinates correspond respectively to $[0001]$, $[\bar1100]$, and $[11\bar20]$ directions. The experimental lattice constants of the hexagonal wurtzite are $a =$ 3.19 \AA, $c =$ 5.19 \AA, and internal displacement between Ga and N sublattices u = 0.377 \cite{SRHMM2000}. With the SW interatomic potential used here, the zero temperature lattice constants are $a =$ 3.19 \AA, $c =$ 5.20 \AA, and u = 0.375. Converting the hexagonal crystal to the smallest orthogonal unit cell, the lattice constants of the unit cell are respectively $a_1 = c =$ 5.2000 \AA, $a_2 = 2\cdot a \cdot cos\left(\pi/6\right) =$ 5.5252 \AA, and $a_3 = a =$ 3.1900 \AA \ in the $x-$, $y-$, and $z-$ directions. For convenience, the system dimension can be represented by the number of cells $n_1$, $n_2$, and $n_3$ in the $x-$, $y-$, and $z-$ directions. In addition to the difference in unit, $n_1$, $n_2$, and $n_3$ always refer to the simulated size whereas $t$, $W$ and $L$ refer to the real size that can become infinite. Two series of sample dimensions, one corresponding to $n_1$ = 100 and the other one corresponding to $n_1$ = 150, were studied at various $n_2$ ranging from 50 to 150 and in addition at $t \rightarrow \infty$ using a fixed $n_3$ = 5. The film scenario with the $(\bar1100)$ film surfaces was simulated using a free boundary condition in the $y-$ direction with surfaces terminated between the larger spacing as will be described in details in the wire case. Such a termination ensures stable surfaces and therefore no surface reconstruction was observed in simulations as supported by the experiment \cite{BLWRURFMBC2009}. The $t \rightarrow \infty$ or $W \rightarrow \infty$ case was simulated by using a periodic boundary condition in the corresponding direction. Note that although the periodic boundary condition is also used in the $x-$ direction, the meaningful dimension in the $x-$ direction for the direct method MD simulations is the spacing between heat source and heat sink. This spacing is not extended by the periodic boundary condition and is always finite.

\subsection{Computational system for wires}

\begin{figure}
\includegraphics[width=4in]{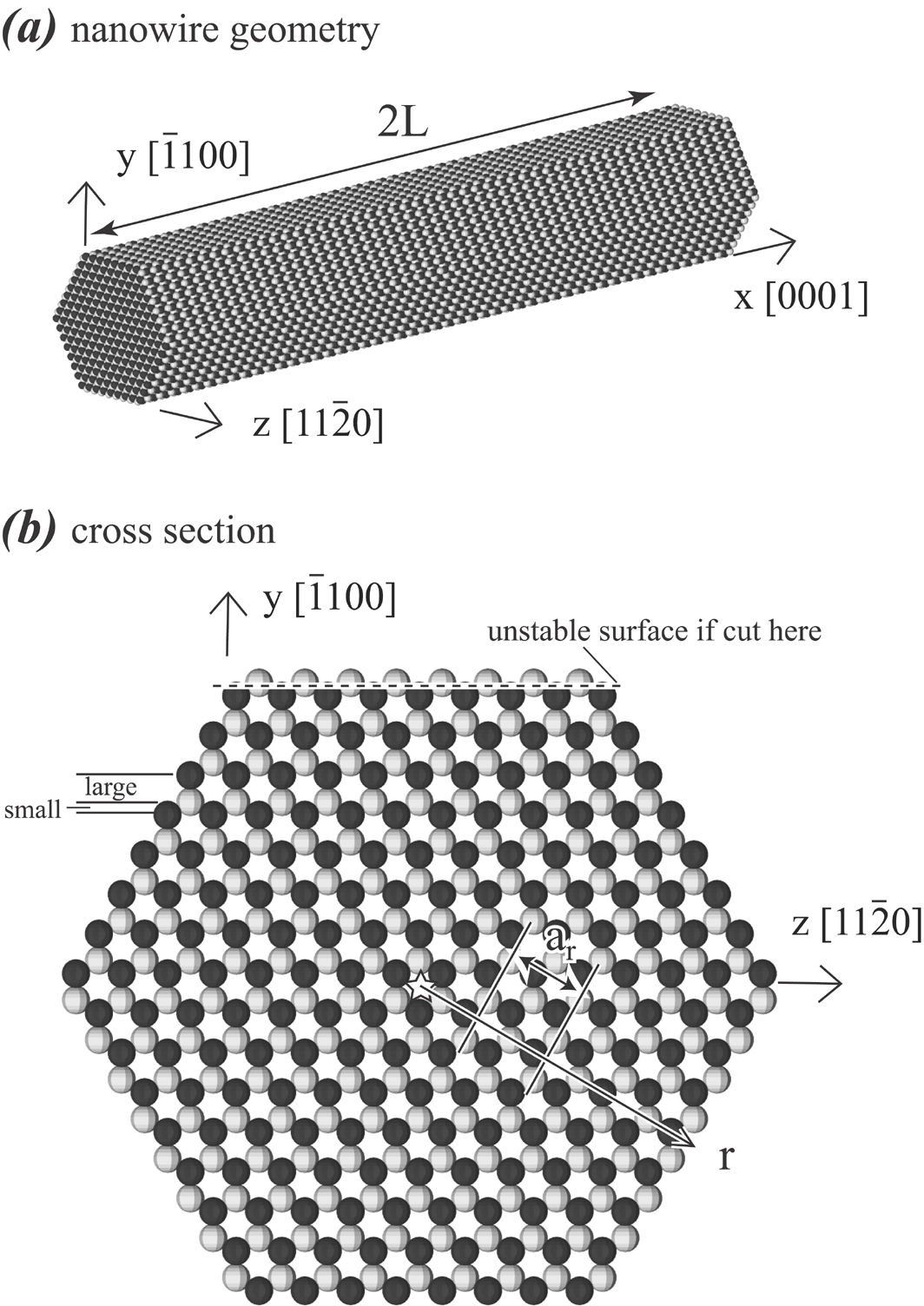}
\caption{Atomistic configuration for wire MD simulations (black and white colors distinguish Ga and N atoms, and the star shows the center of the wire cross section). 
\label{wire MD model}}
\end{figure}

Unlike the circular wire assumed in Fig. \ref{wire model}, the GaN wires observed in experiments are often hexagonal with $\left<0001\right>$ axis and $\{\bar1100\}$ facets. The computational system used for simulations of such wires is shown in Fig. \ref{wire MD model}(a). The crystal orientations are the same as those used for the film simulations, except that atoms beyond specified $\{\bar1100\}$ surfaces are removed. As in Fig. \ref{wire model}, the system is assumed to have a finite radius $r$, which is defined as the minimum distance between the center of the wire and the surface, Fig. \ref{wire MD model}(b). Again the system dimension in the x- direction is assumed to be $2L$ rather than $L$ to facilitate the periodic boundary condition. 

The cross section of the hexagonal wire is examined in details in Fig. \ref{wire MD model}(b). It can be seen that the $\{\bar1100\}$ atomic planes have two different spacings. We found that if the surface is terminated between the small spacing, such as that shown by the dash line, then the surface is composed of many dangling bonds resulting in high energy and unstable configurations. Hence, the surfaces of our wires are always terminated between the large spacing. It can also be seen that the smallest repeatable spacing of the $\{\bar1100\}$ planes is $a_r = a \cdot cos\left(\pi/6\right) =$ 2.7616 \AA. For convenience, the system dimension in the $x-$ and the radial directions can be respectively represented by number of cells $n_1$ and $n_r$ (in unit of $a_r$). A matrix of dimensions with the longitudinal dimension $n_1$ ranging from 100 to 300 and the radial dimension $n_r$ ranging from 4 to 12 was explored. Here free boundary conditions were used in the $y-$ and $z-$ directions and the periodic boundary condition used in the $x-$ direction.

\subsection{Heat transport simulation algorithm}

The thermal transport MD simulations were performed under a constant number of atoms, constant system volume, and constant system energy (NVE) condition using a time step size of $\Delta t = 1 fs$. To accurately account for the effect of thermal expansion and eliminate the errors due to statistical fluctuation of the simulated temperature, the following steps were used to create the initial crystal. First, a crystal was created by assigning atom positions according to the prescribed crystal lattice and the known lattice constants at zero temperature. A molecular dynamics simulation in the constant atom number, (zero) pressure, and temperature (NPT) ensemble was subsequently performed for a total of $20 ps$ period. The desired simulated temperature was achieved using the velocity rescaling method. After discarding the first $10 ps$ simulation to allow the system to reach a steady state, the average crystal sizes and average total (kinetic and potential) system energy were then calculated for the remaining $10 ps$. We then created another crystal using the average sizes obtained at the finite temperature. We can also calculate the potential energy of this newly created crystal by simply performing an energy calculation simulation. The difference between the average total system energy obtained from the NPT run and the potential energy of this crystal prescribes exactly the amount of the kinetic energy that needs to be added in order for this crystal to exhibit the same total average energy over the subsequent long constant energy thermal transport simulation. We added precisely this amount of kinetic energy into the system by first assigning velocities to atoms according to Boltzmann distribution and then rescaling the velocities under the zero total linear momentum condition \cite{IH1994,JJ1999,WZS2008}. The thermal transport simulation is started immediately without the conventional long NPT or NVT simulation to establish the initial temperature. The advantage of this approach is that once steady-state is reached, the average temperature of the system matches exactly the desired temperature. In practice, we found that the difference between the average temperature in the middle of the heat source and heat sink (where the thermal conductivity was calculated) and the desired temperature is well below 1 K (often  near 0.01 K or less). This method has the same accuracy level as the ``doubling temperature'' method applied previously \cite{ZAJGS2009}. It is more general because the initial crystal is not required to be in the minimum potential energy configuration, which may not be easily determined when the system includes surfaces.

The direct method requires the creation of a heat source and a heat sink. As shown in Fig. \ref{thin film MD model}, the heat source corresponds to the red region at the far left of the cell, and the heat sink is the blue region near the middle of the cell. During simulations, the size and location of the source and sink regions are specified. With appropriate choices of system length, source and sink region width, and locations of different regions, we ensured that the source and sink regions were geometrically identical, and that the left side of the sink (or source) region was exactly symmetric to its right side, up to another source (or sink) region (which may be its periodic image). Here the width of the source or sink region is around 40 \AA. It can be seen from Fig. \ref{thin film MD model} that the heat source at the left side has an image at the right side under the periodic boundary condition in the x- direction. Consequently, even the length of our system is $2L$, the spacing between the source and the sink is still $L$ as shown in Fig. \ref{film model}.

The constant flux \cite{IH1994,SP2001,JJ1999,SPK2002,SPK2004,YCSS2004} method was used to create the temperature gradient. In this method, a constant amount of energy is added to the hot region and exactly the same amount of energy is removed from the cold region at each MD time step using velocity rescaling (while preserving linear momentum). To ensure that the high and the low temperatures are reasonable and consistent in different runs, the heat flux has been adjusted within 0.00035 to 0.0010 $eV/(ps \cdot \AA^2)$ to give a consistent high-low temperature difference (say 5 - 10 K). To generate extremely accurate results, the first 0.4 ns simulation was discarded to allow the system to reach a steady state, and the remaining duration of simulations was chosen to be at least 11 ns and many reached over 20 ns. To compute the temperature profile, the system dimension in the $x-$ direction is divided into a grid. The temperature of each of the grid cells was averaged over the remaining time of simulations. The temperature profile and the input heat flux were used to calculate the thermal conductivity using Fourier's Law, Eq. (\ref{kappa 1}). To estimate the statistical error of the calculated thermal conductivity, the total averaging time was divided into 20 subsections and thermal resistivity (or conductivity) was calculated for each of the subsections. From these data, analysis of statistical error was performed. For more details on the procedure please see \cite{ZAJGS2009}.

It should be noted that under the periodic boundary condition in the x- direction, the observed dependence of thermal conductivity upon the system length $L$ comes primarily from the scattering of the interfaces at the hot and the cold regions. The functional dependence on $L$ can still be well described by Eqs. (\ref{kappa 3}) and (\ref{kappa 4}), albeit $\delta_i$ should be viewed as an interface scattering parameter rather than surface scattering parameter. Because our analysis extrapolates the MD data to an $L \rightarrow \infty$ limit (i.e., true film and true wire), the interface approximation will not affect the results.

\section{Results}

\subsection{Film thermal conductivity as a function of film thickness}

\begin{figure}
\includegraphics[width=6in]{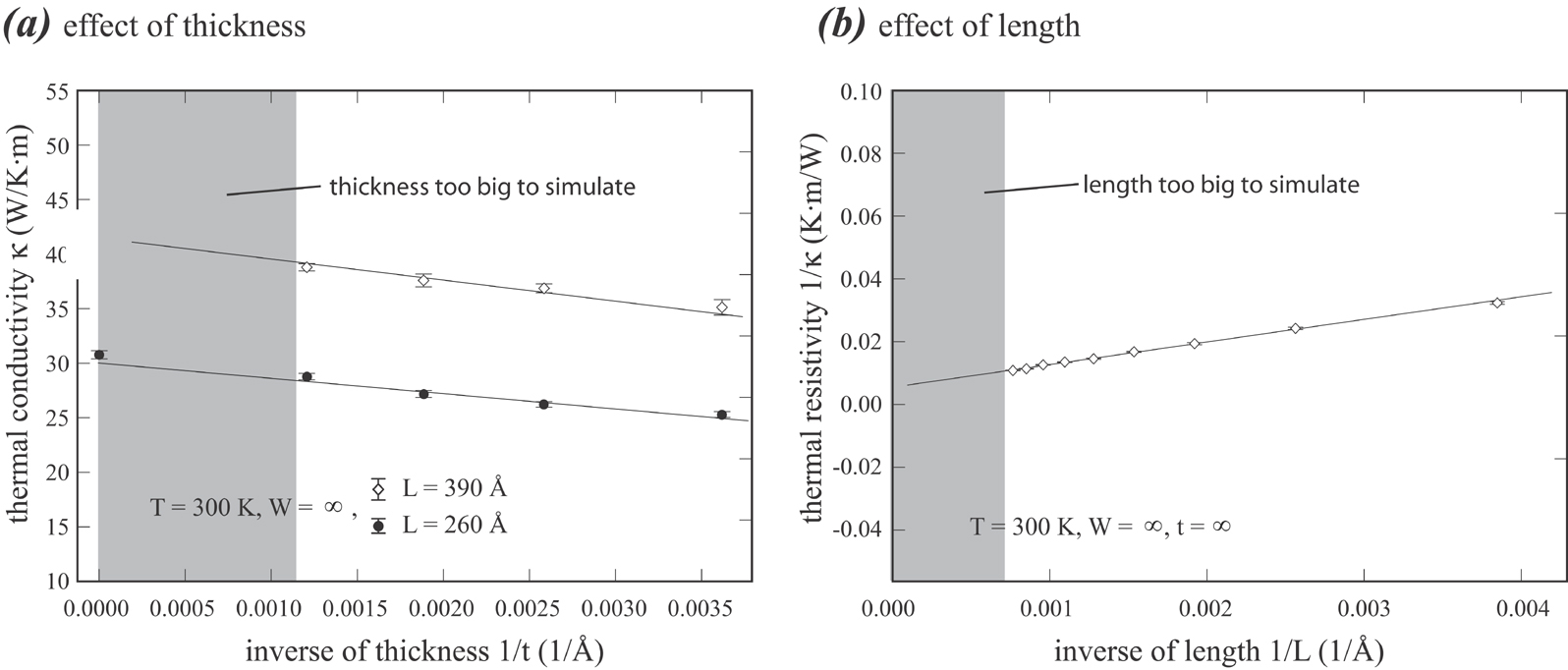}
\caption{Thermal conductivity data obtained for film at 300 K.
\label{300 K film data}}
\end{figure}

\begin{figure}
\includegraphics[width=6in]{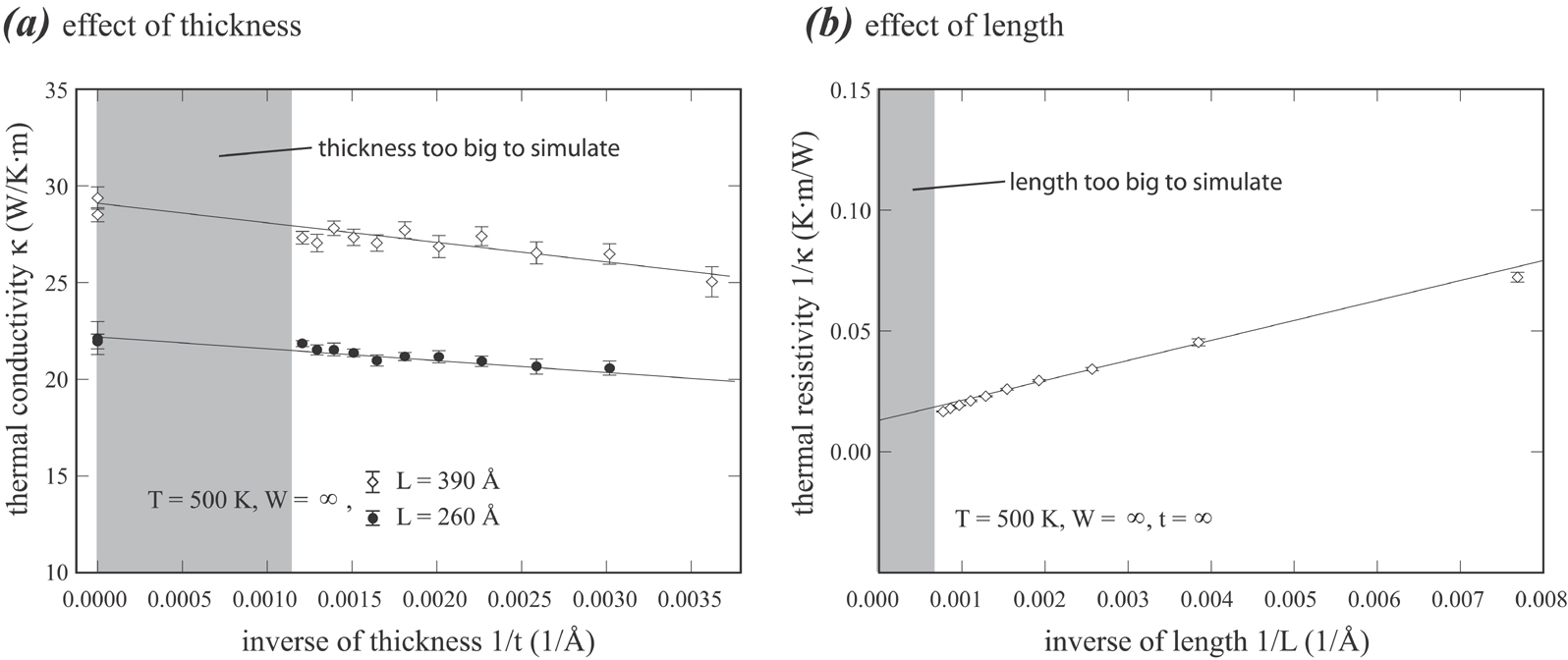}
\caption{Thermal conductivity data obtained for film at 500 K.
\label{500 K film data}}
\end{figure}

Systematic MD simulations were performed to derive thermal conductivity of film as a function of film thickness at two temperatures of T = 300 K and T = 500 K. As the Debye temperature was estimated to be in the range 350-600 K \cite{S1973,MAG1998,DPWJP2006}, the system is expected to behave classically especially at 500 K. Although 300 K is at the lower bound of the estimated Debye temperature range, it is chosen for study because the low temperature data has less thermal fluctuation and therefore can provide stronger model validation. At each temperature, two series of MD simulations corresponding respectively to $2L \approx 520 \AA$~($n_1 = 100$) and $2L \approx 780 \AA$~($n_1 = 150$), were performed at different thickness $t$ but fixed width $W \rightarrow \infty$ as described in the above. Previous MD simulations have determined the thermal conductivities at various length $L$ but a fixed thickness $t \rightarrow \infty$ and a fixed width $W \rightarrow \infty$ \cite{ZAJGS2009}. Both present and previous data are collectively used to fit Eq. (\ref{kappa 3}) using a chosen value of $d$, and the results of the fitted parameters are shown in Table \ref{parameters}. Both the MD data and the fitted curves are shown in Figs. \ref{300 K film data} and \ref{500 K film data} for the 300 K and 500 K temperatures respectively. Eq. (\ref{kappa 3}) indicates that at a fixed $L$, thermal conductivity $\kappa$ is a linear function of $1/t$, and at a fixed $t$, the inverse of thermal conductivity $1/\kappa$ is approximately a linear function of $1/L$. Hence, Figs. \ref{300 K film data}(a) and \ref{500 K film data}(a) show the $\kappa$ vs. $1/t$ plots at fixed $L$ whereas Figs. \ref{300 K film data}(b) and \ref{500 K film data}(b) show the $1/\kappa$ vs. $1/L$ plots at fixed $t$. It can be seen that the linear relationships predicted by Eq. (\ref{kappa 3}) is strikingly reproduced by the MD data, and the agreement between the fitted lines and the MD data is excellent using a single set of parameters ($d$, $\kappa_{0,c}$, $\kappa_{1,c}$, $\delta_0$, and $\delta_1$) for both the thickness and the length functions.
\begin{table}
\caption{Parameters describing the general scaling law.}
\begin{ruledtabular}
\begin{tabular}{rrrrrrr} 
structure&T$~(K)$&$d~(\AA)$&$\kappa_{0,c} (W/K \cdot m)$&$\kappa_{1,c}~(W/K \cdot m)$&$\delta_0~(\AA)$&$\delta_1~(\AA)$\\\hline 
film&$300$&$138.13$&$178.38$&$151.65$&$1288.10$&$1329.65$\\\hline 
film&$500$&$138.13$&$75.42$&$56.16$&$623.38$&$471.74$\\\hline
wire&$500$&$11.05$&$75.42$&$47.30$&$615.51$&$1324.45$\\\hline
wire&$500$&$22.09$&$75.42$&$50.30$&$604.71$&$795.87$\\
\end{tabular}
\end{ruledtabular}
\label{parameters}
\end{table}

In Fig. \ref{300 K film data} and \ref{500 K film data}, the shaded area indicate the length-scale no-man's land, meaning that the system dimension is too large to be directly simulated using the MD methods. Fig. \ref{300 K film data}(b) and \ref{500 K film data}(b) show that the linear scaling law can be used to predict thermal conductivity in the shaded region through extrapolation based upon the MD data obtained in a small dimension range. Although the linear extrapolation is expected to produce reliable results, the accuracy of the extrapolated values is difficult to confirm directly. In sharp contrast, Fig. \ref{300 K film data}(a) and \ref{500 K film data}(a) show that because the thermal conductivity at an infinite thickness $t$ can be obtained in the MD simulation by using periodic boundary condition and because the linear relationship extends to this value at $t \rightarrow \infty$, an extremely reliable interpolation can be used to predict thermal conductivity at any thickness dimension. Now we can see that the satisfaction of the linear relationships by the MD data and the extrapolation of the linear relation to the bulk limit can be used to determine if the selected $d$ value is sufficiently large.

\begin{figure}
\includegraphics[width=6in]{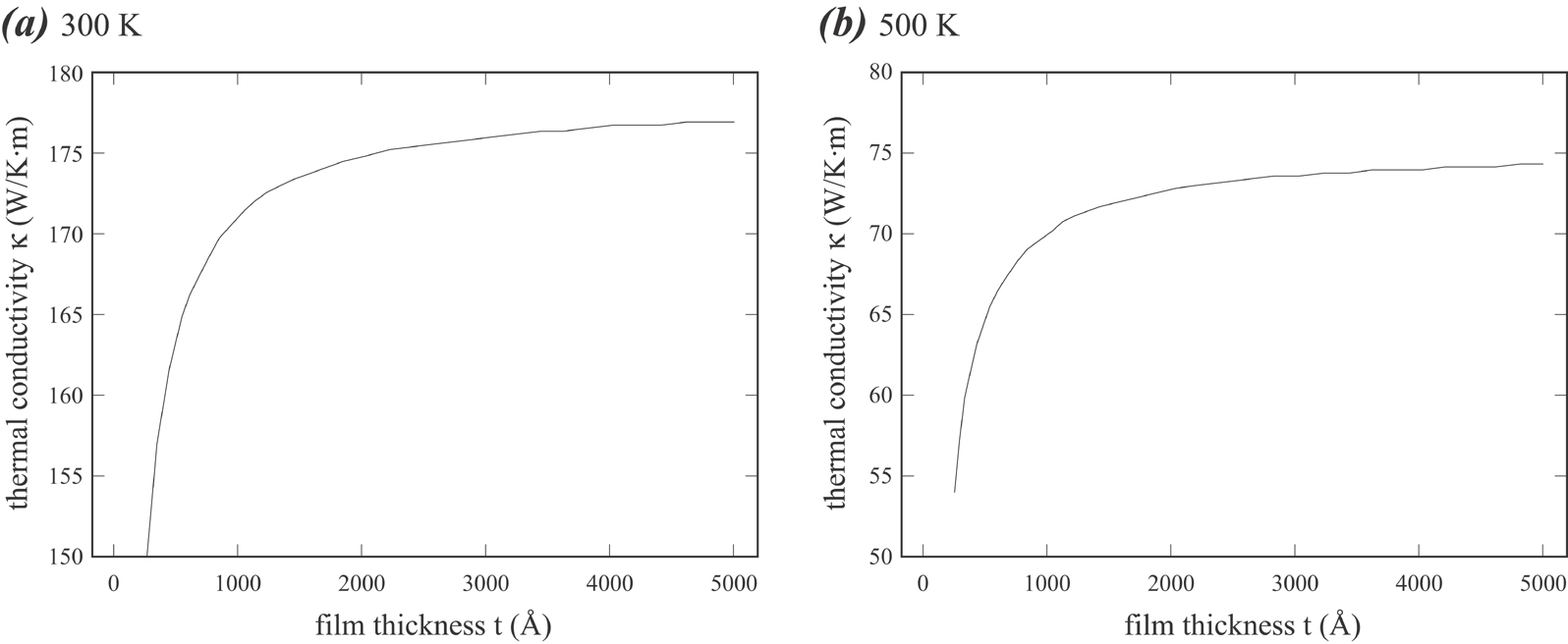}
\caption{Predicted thermal conductivity of film as a function of film thickness.
\label{kappa vs film thickness}}
\end{figure}

For true films, we set $L \rightarrow \infty$. Using the parameters listed in Table \ref{parameters} and Eq. (\ref{kappa 3}), thermal conductivity of film was calculated as a function of film thickness, and the results obtained at 300 K and 500 K temperatures are shown respectively in Figs. \ref{kappa vs film thickness}(a) and \ref{kappa vs film thickness}(b).

\begin{figure}
\subfigure{\includegraphics[width=4in]{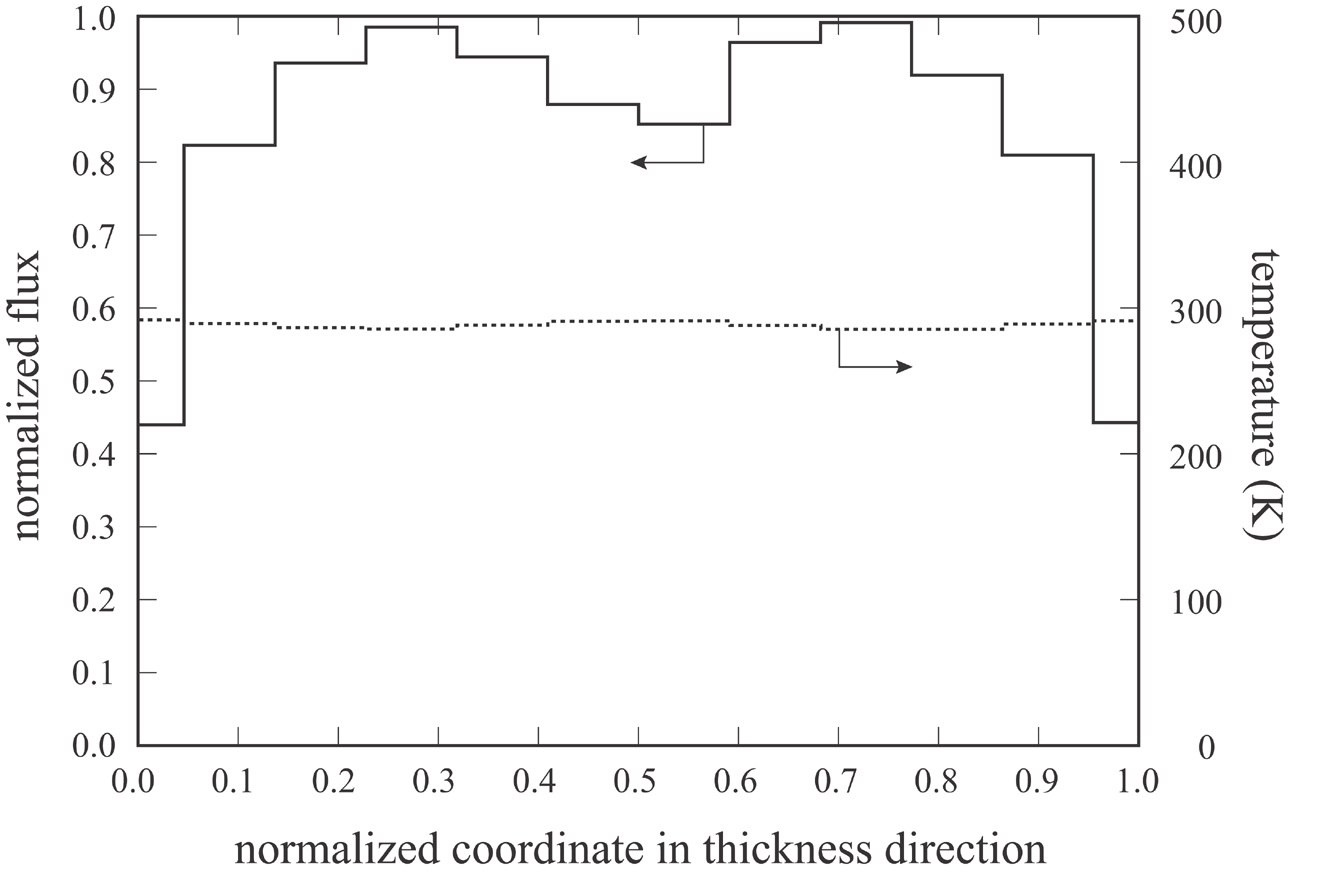}}
\caption{Flux and temperature profiles along the thickness of a thin film.
\label{film_profile}}
\end{figure}

To confirm our hypothesis that the boundary scattering region is confined to the surface, we calculated the heat flux of a thin film in slabs parallel the surface of the film. The calculated relative flux and temperature profiles are shown in Fig. \ref{film_profile}. Clearly it can be seen that flux remains nearly constant in the interior but is degraded near surfaces. Temperature, on the other hand, is constant across the entire thickness of the sample. This strongly validates that the conductivity in the interior is a constant whereas it is reduced at surfaces. This phenomenon is unknown in the past. It indicates that although the thermal conductivity at small scale can be thought to depend on the phonon mean free path, it can still be manifested through the core-shell phenomenon as assumed in our model. This accounts for why our scaling law agrees well with the effects deduced from considering only the phonon mean free path \cite{SPK2002}.

\subsection{Wire thermal conductivity as a function of wire radius}

\begin{figure}
\includegraphics[width=6in]{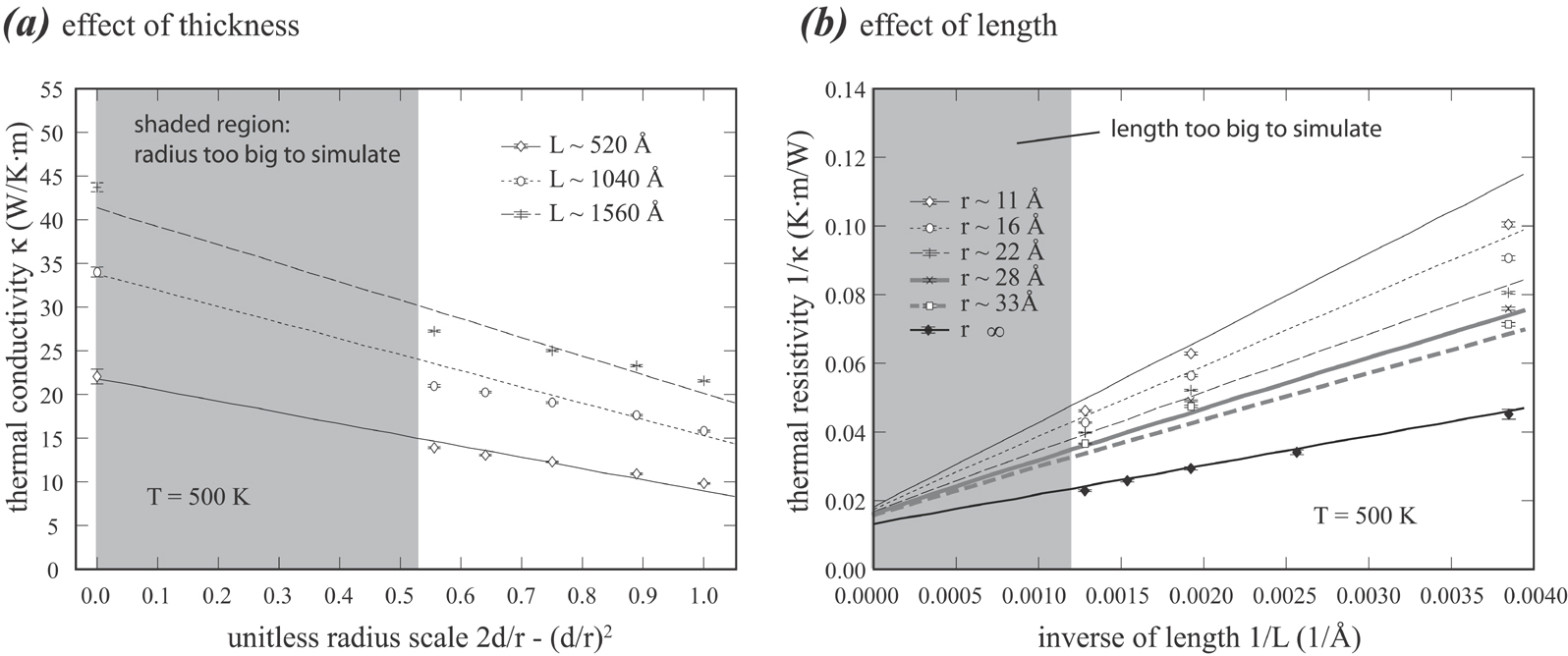}
\caption{Thermal conductivity data obtained for wire at 500 K, $d = 11.05 \AA$.
\label{500 K wire data small d}}
\end{figure}

Having verified the film case at both 300 K and 500 K, MD simulations were performed to study thermal transport for wires at 500 K. Only one temperature is studied due to the expense of the calculations. 500 K was chosen because it is likely to be above the Debye temperature. To deduce all the parameters, simulations were carried out using a matrix of three sample length of $2L \approx 520 \AA$~($n_1 = 100$), $1040 \AA$~($n_1 = 200$), and $1560 \AA$~($n_1 = 300$) and four sample radius of $r \approx 11 \AA$~($n_r = 4$), $16 \AA$~($n_r = 6$), $22 \AA$~($n_r = 8$), and $33 \AA$~($n_r = 12$). Eq. (\ref{kappa 4}) indicates that at a given length $L$, $\kappa$ is a linear function of $2d/r - (d/r)^2$ and at a given radius $r$, $1/\kappa$ is approximately a linear function of $1/L$. Here we explore two different chosen $d$ values, $d = 11.05 \AA$ and $d = 22.09 \AA$. Using $d = 11.05 \AA$ (which is the largest $d$ that enables all MD data to satisfy the geometry condition of the scaling model), $\kappa$ vs $2d/r - (d/r)^2$ curves at different $L$ are shown in Fig. \ref{500 K wire data small d}(a) and $1/\kappa$ vs $1/L$ curves at different $r$ are shown in Fig. \ref{500 K wire data small d}(b). In addition, the $\kappa$ values obtained at infinite cross section dimension (i.e., $r \rightarrow \infty$) obtained in the previous work \cite{ZAJGS2009} are included in Fig. \ref{500 K wire data small d}. In Fig. \ref{500 K wire data small d}, the lines are calculated using Eq. (\ref{kappa 4}) with fitted parameters displayed in Table \ref{parameters}. It can be seen that Fig. \ref{500 K wire data small d} exhibits some linear relationships predicted by Eq. (\ref{kappa 4}). However, the overall match between the MD data and the model prediction is not great. Most seriously, Fig. \ref{500 K wire data small d}(a) indicates that a linear regression using merely the data points at large $2d/r - (d/r)^2$ values would not closely extrapolate to the data point at $2d/r - (d/r)^2 = 0$ (i.e., $r \rightarrow \infty$), and Fig. \ref{500 K wire data small d}(b) shows significant deviation of the predicted curves from the data points.

\begin{figure}
\includegraphics[width=6in]{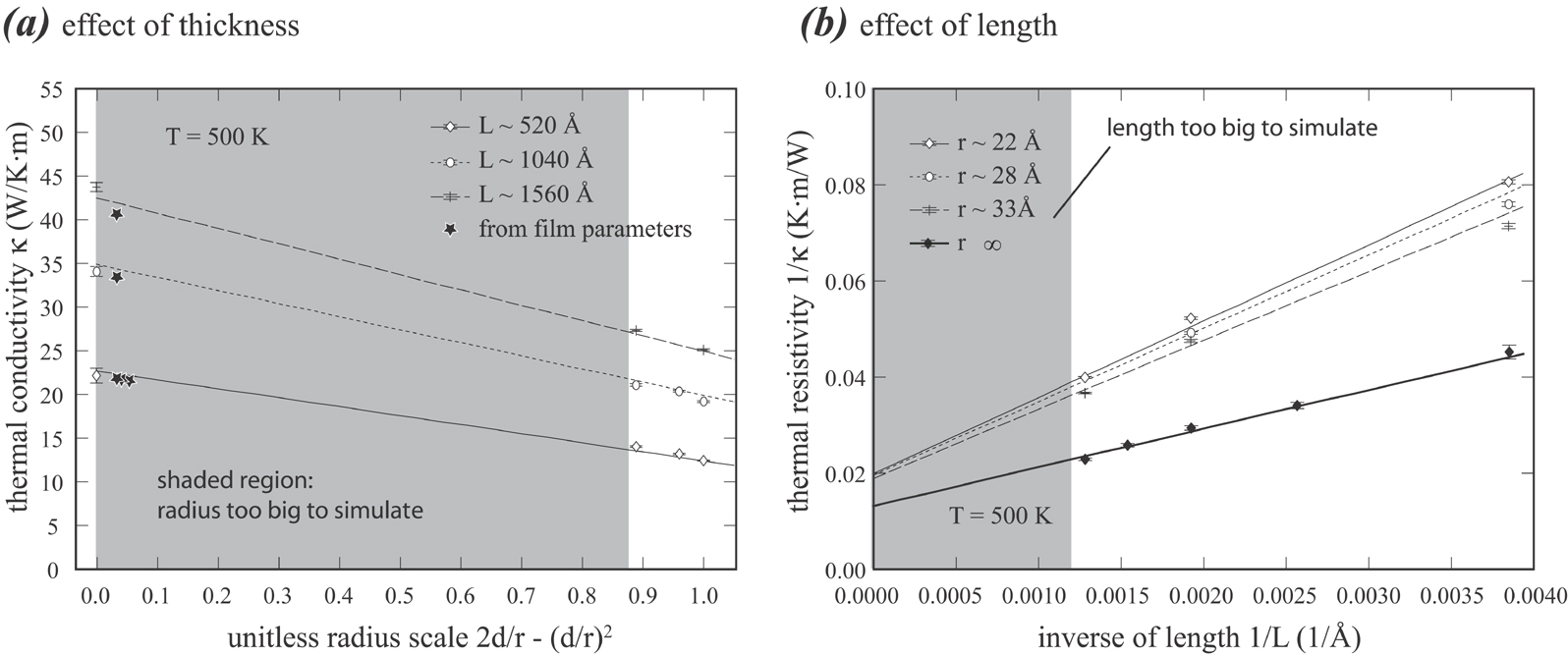}
\caption{Thermal conductivity data obtained for wire at 500 K, $d = 22.09 \AA$.
\label{500 K wire data big d}}
\end{figure}

The problems exhibited by Fig. \ref{500 K wire data small d} is inherently related to an important condition: the parameter $d$ must be chosen to be sufficiently large to subsume the surface scattering effect in order for Eq. (\ref{kappa 4}) to be valid. To explore this further, we increased $d$ to $22.09~\AA$. This disqualified the data obtained at small radii of $r \approx 11~\AA$~and $r \approx 16~\AA$. The remaining data were fitted to Eq. (\ref{kappa 4}), and the parameters thus obtained are included in Table \ref{parameters}. With the increased value of $d$, the results similar to those in Figs. \ref{500 K wire data small d}(a) and \ref{500 K wire data small d}(b) are recalculated and are presented in Figs. \ref{500 K wire data big d}(a) and \ref{500 K wire data big d}(b) respectively (note that the scales in the horizontal axis are different between Figs. \ref{500 K wire data small d} and \ref{500 K wire data big d} due to the use of different $d$ values). Clearly, a significant improvement is achieved. In particular, Fig. \ref{500 K wire data big d}(a) indicates that the data points obtained at large $2d/r - (d/r)^2$ values can be linearly connected to accurately extrapolate to the point at $2d/r - (d/r)^2$ = 0, and Fig. \ref{500 K wire data big d}(b) shows an improved agreement between data points and the predicted curves.

The value of $d$ used in the wire case is constrained by the radius $r$ used in the MD simulations. Unlike the film case where the thickness $t$ is independent of the width $W$, increasing the radius in the wire case would increase the system dimensions in both y- and z- directions. As a result, the magnitude of the radius is more severely constrained by the computational expense. In general, current computer resources may not permit the use of radius significantly above $n_r = 12$. The $d$ value permitted by this radius is still small and significant further improvement over the one seen in Fig. \ref{500 K wire data big d} is likely to be achieved if $d$ can be substantially larger. To improve the results, we explored an alternative method by again using the scaling theory. Film simulations described in the previous section resulted in the determination of five parameters $\kappa_{0,c}$, $\kappa_{1,c}$, $\delta_0$, $\delta_1$, and $d$. These five parameters have physical meanings and are invariant in the wire configurations. As a result, we can directly use these parameters and Eq. (\ref{kappa 4}) to calculate wire thermal conductivity. Note that due to the relatively low computation cost, the $d$ value used in the film case reaches $138~\AA$, Table \ref{parameters}. As a result, significantly more accurate results can be expected. Some results of this calculations at a few different radii $r \ge d = 138.13~\AA$ are included in Fig. \ref{500 K wire data big d}(b) using stars. It can be seen that although the predicted results from film simulations using big $d = 138~\AA$~and wire simulations using small $d = 22.09 \AA$~are different, the difference is relatively small at least for GaN. Note that while the film parameters based upon a big $d = 138~\AA$~ give better results, they cannot be used to calculate thermal conductivity at $r < d = 138~\AA$.  

\begin{figure}
\includegraphics[width=4in]{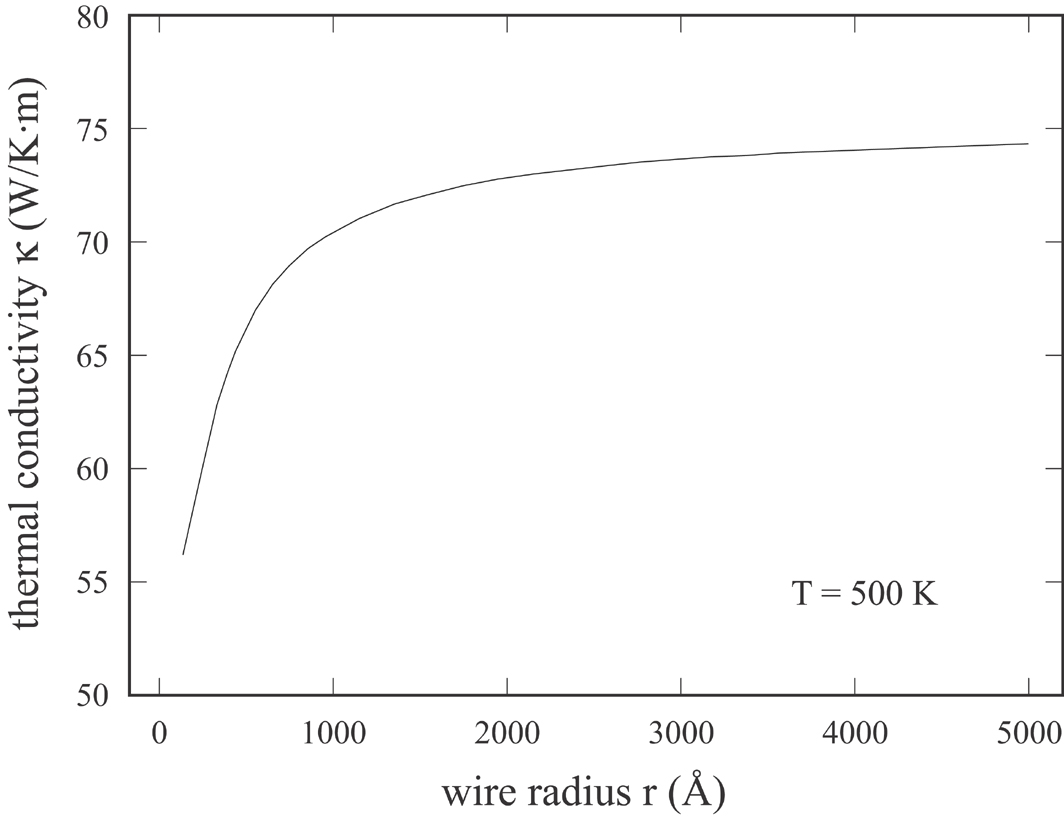}
\caption{Predicted thermal conductivity of wire as a function of wire radius.
\label{kappa vs wire radius}}
\end{figure}

For true wires, we set $L \rightarrow \infty$. The film parameters and Eq. (\ref{kappa 4}) were used to predict wire thermal conductivity as a function of radius. The results are shown in Fig. \ref{kappa vs wire radius}.

\section{Discussion}

\subsection{Generalized scaling law}

The scaling law described in the above can be generalized to any prismatic wire with arbitrary cross-section. Approximated as coarse-grained resistance network with temperature being a function of $x$-coordinate only, the cross section of such a wire can be broken up into a finite number of distinct areas $A_i$, with one area $A_0$ completely in the interior of the wire and all the others $i>0$ touching the boundary of the wire. The total cross-sectional area is then $A = \sum_i A_i$. The total flux $J$ is partitioned such that
\begin{equation}
A J = \sum_i J_i A_i  = \sum_i \kappa_{i,j} \frac{\Delta T_j}{\Delta x_j} A_i  
\end{equation}
where the second equality is obtained from Fourier's law, and $\kappa_{i,j}$ is the conductivity of the $i$-th area and the $j$-th section. Here we use $j$ = -1 and 1 to indicate the two end sections and $j$ = 0 to indicate the middle section. It is assumed that the two end sections are similar so that $\kappa_{i,-1} = \kappa_{i,1} = \kappa_{i,\pm{1}}$ and $\Delta T_{-1} = \Delta T_{1} = \Delta T_{\pm{1}}$. Under the geometry conditions that $\Delta x_{-1}= \Delta x_{1} = d$ and $\Delta x_{-1} + \Delta x_{0} + \Delta x_{1} = L$, we have 
\begin{equation}
A J = \sum_i \kappa_{i,\pm{1}} \frac{\Delta T_{\pm{1}}}{d} A_i
    = \sum_i \kappa_{i,0} \frac{\Delta T_0}{L-2d} A_i
\label{A J 1}
\end{equation}
Using the continuous temperature field condition $\Delta T = 2 \Delta T_{\pm{1}} + \Delta T_{0}$, Eq. (\ref{A J 1}) can be written as
\begin{equation}
A J \left( \frac{2 d}{\sum_i \kappa_{i,\pm{1}} A_i} + \frac{L-2d}{\sum_i \kappa_{i,0} A_i} \right) = \Delta T
\end{equation}
The constituent areas can be classified into three types: corners, boundaries, and the interior. Each relates to the overall radius $R$ of the wire. In a self-similar fashion, the interior scales with $R^2$; the boundaries scale with $R$; and the corners are essentially constant. Using geometry-specific constants $C_i$, the areas of different types can be expressed as $A_i/A = C_i/R^2$ for the corners, $A_i/A = C_i/R$ for the boundaries, and $A_0/A = 1 - \sum_{i>0} A_i/A$ for the interior. The overall thermal conductivity of the wire is then
\begin{eqnarray}
\kappa &=& \frac{J L}{\Delta T} \nonumber \\ &=& \left[ 
\frac{2 d/L}{\kappa_{0,\pm{1}} (1 - 1/R \sum_{i\in \mathcal{B}} C_i - 1/R^2 \sum_{i\in \mathcal{V}} C_i) + 1/R \sum_{i\in \mathcal{B}} \kappa_{i,\pm{1}} C_i + 1/R^2 \sum_{i\in \mathcal{V}} \kappa_{i,\pm{1}} C_i} \right. \nonumber \\ &+& \left.
\frac{1-2d/L}{\kappa_{0,0} (1 - 1/R \sum_{i\in \mathcal{B}} C_i - 1/R^2 \sum_{i\in \mathcal{V}} C_i) + 1/R \sum_{i\in \mathcal{B}} \kappa_{i,0} C_i + 1/R^2 \sum_{i\in \mathcal{V}} \kappa_{i,0} C_i} \right]^{-1}
\label{generalized kappa 1}
\end{eqnarray}
where $\mathcal{B}$ and $\mathcal{V}$ represent the sets of boundaries and corners respectively, $L$ and $R$ are input sample dimensions, and the rest of the parameters, i.e. $d$, $C_i$, and $\kappa_{i,j}$, need to be fitted. It can be seen that Eq. (\ref{generalized kappa 1}) correctly reduces to the bulk thermal conductivity value $\kappa_{0,0}$ as $L\rightarrow \infty$ and $R \rightarrow \infty$. Practical nanowires usually have symmetric cross section, thus it is possible to reduce the number of independent parameters. For an equilateral triangular prismatic wire, for example, the free parameters would include $\kappa_{0,\pm{1}}$ and $\kappa_{0,0}$ for the core, $\kappa_{1,\pm{1}}$ and $\kappa_{1,0}$ for the boundaries, and $\kappa_{2,\pm{1}}$ and $\kappa_{2,0}$ for the vertices in addition to $d$, $C_1$, and $C_2$. Eq. (\ref{generalized kappa 1}) then becomes
\begin{eqnarray}
\kappa &=& \left[ \frac{2 d/L}{\kappa_{0,\pm{1}} (1 - C_1/R - C_2/R^2) + C_1 \cdot \kappa_{1,\pm{1}}/R  + C_2 \cdot \kappa_{2,\pm{1}}/R^2} \right. \nonumber \\ &+& \left. \frac{2 d/L}{\kappa_{0,0}  (1 - C_1/R - C_2/R^2) + C_1 \cdot \kappa_{1,0}/R + C_2 \cdot \kappa_{2,0}/R^2} \right]^{-1}
\label{generalized kappa 2}
\end{eqnarray}

\subsection{Effect of periodic boundary conditions}

\begin{figure}
\includegraphics[width=4in]{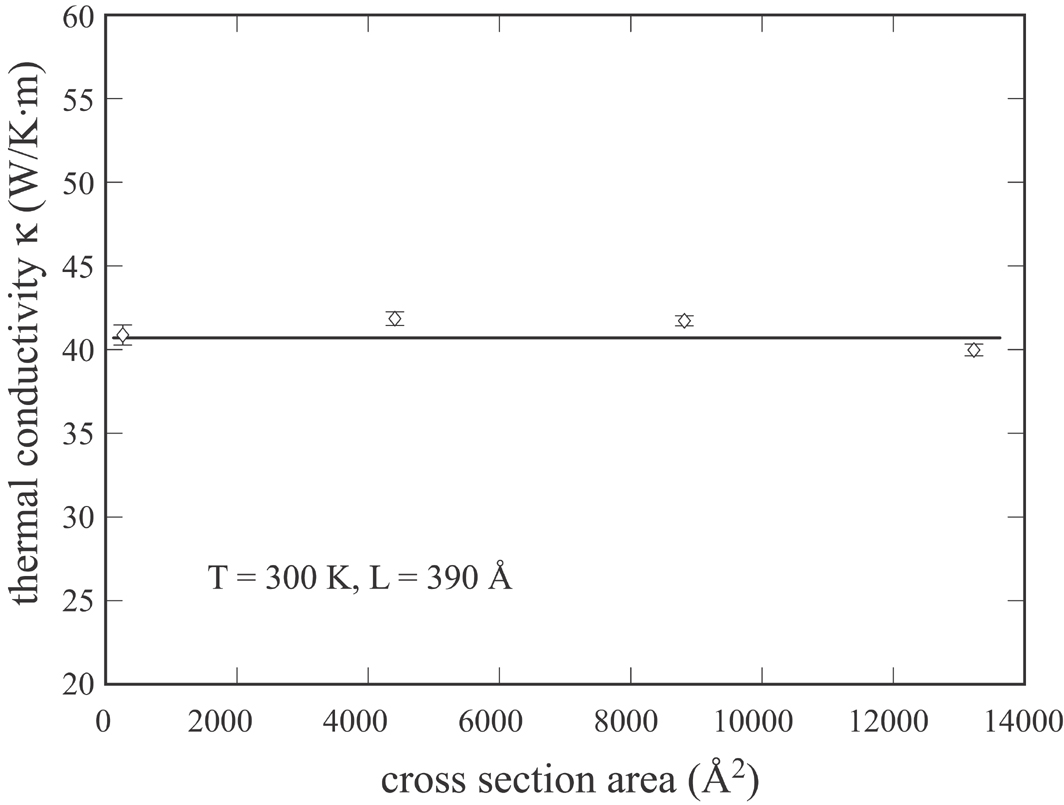}
\caption{Effect of cross section area on thermal conductivity under the periodic boundary condition.
\label{area effect}}
\end{figure}

The use of periodic boundary condition eliminates surfaces. Strictly speaking, however, there is still an interface between the periodic image that is not exactly the same as if there were no such an image, e.g. phonon wavelength is restricted by the size of the periodic cell. The effect of this interface has been explored previously by using the periodic boundary conditions in the two cross section directions with different simulated cross section areas \cite{ZAJGS2009}. The results showed that the periodic interfaces parallel to the heat flow have no significant effect on thermal conductivity \cite{ZAJGS2009}. As a result, periodic boundary condition was thought to be able to extend the system dimension to infinite. The past study, however, only examined the simulated cross section area between 180 $\AA^2$ and 890 $\AA^2$. In the present work, the cross section area reached as high as about 13250 $\AA^2$. Hence, we re-examine the effect of cross section area under periodic boundary conditions. Thermal conductivities were calculated at 300 K using periodic boundary conditions in both y- and z- directions with fixed simulated dimension of $n_1$ = 150 ($2L \approx 780 \AA$), $n_3$ = 5 ($\approx 16 \AA$) and various thickness of $n_2$ between 3 ($\approx 16 \AA$) and 150 ($\approx 830 \AA$). The results are shown in Fig. \ref{area effect}. Fig. \ref{area effect} confirms that when the periodic boundary condition is used to extend the cross section dimension to infinity, the magnitude of the simulated dimension does not affect the thermal conductivity in a wide range of cross section between $260 \AA^2$ and $13220 \AA^2$. Furthermore, since we only varied the thickness $t$, our result provides a strong evidence that the aspect ratio $t/W$ also does not affect the thermal conductivity under the periodic boundary condition.

\subsection{Effect of surface stress}

\begin{figure}
\includegraphics[width=4in]{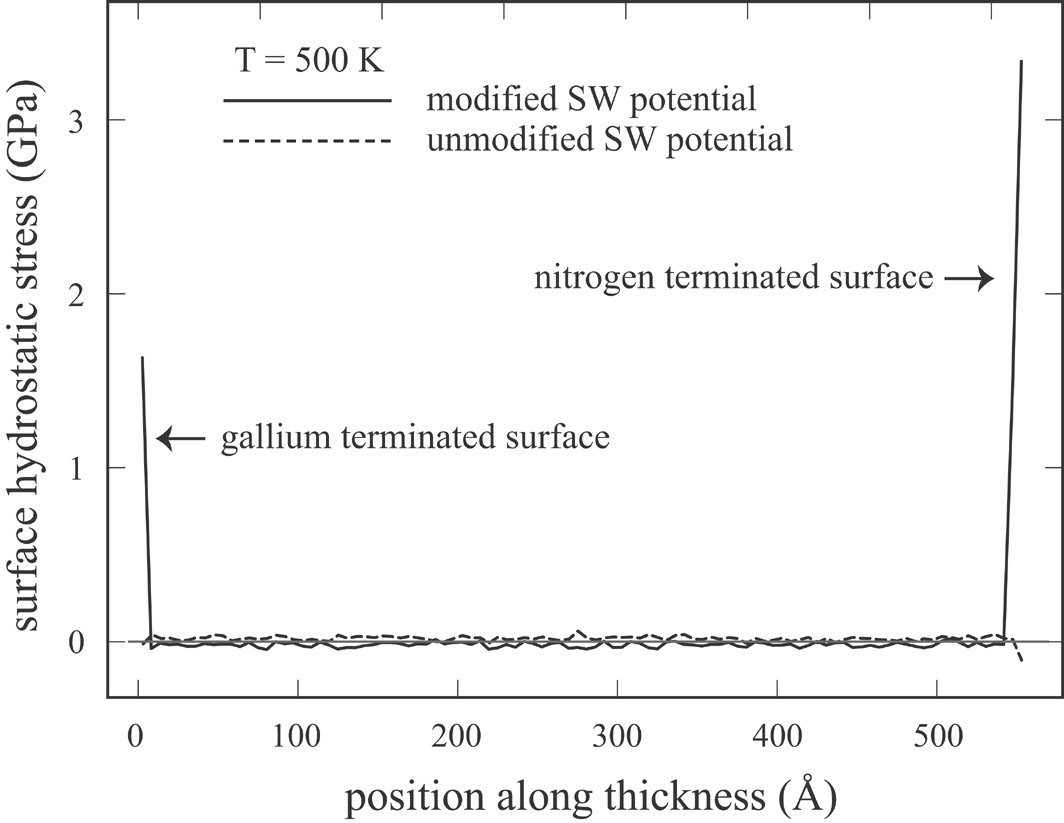}
\caption{Stress profile along the film thickness.
\label{stress profile}}
\end{figure}

The simulations described above are based upon a Stillinger-Weber potential. An essential feature of SW potential is that its pairwise functions decay to negligible values within the second nearest neighbor distance of atoms. A problem with the nearest-neighbor potentials is that they predict zero surface stresses \cite{BHT1992}. The neglect of surface stresses may alter the thermal conductivities of nanostructures. We have also explored the use of alternative potentials such as the Tersoff GaN potential \cite{NAEN2003} in our studies. Unfortunately, we discovered that the Tersoff potential severely underestimates the thermal conductivity as compared with the experimental measurement. With the SW potential clearly the better choice for the thermal transport simulations, we carefully assess the surface stress effect. This was done by modifying the SW potential to artificially create a surface stress. All the pairwise functions used in the SW potential are expressed in terms of the scaled atomic spacing as $F(R/\sigma)$ \cite{BS2002,BS2006}, where $R$ is the atomic spacing and $\sigma$ is a scaling factor. As a result, we scaled the parameter $\sigma$ by 0.98 to reduce the equilibrium bond length by 2 \%. We then used the modified potential for the two atomic layers of atoms on the nanostructure surfaces and the unmodified potential for the remaining atoms. This effectively created a surface tensile stress. This is a good approximation because it realistically captures the surface stress in the subsurface region despite the variation of atomic interaction of the surface atoms. This modified scheme of interatomic potential was then used to perform a selected case of film simulation at a temperature of 500 K, a sample length $n_1$ = 150 ($2L \approx 780 \AA$), a width $n_3$ = 5 ($16 \AA$), and a thickness $n_2$ = 100 ($\approx 550 \AA$). To understand the surface stress of the sample, the system was first relaxed using molecular statics energy minimization simulation at a constant volume determined from an NPT MD simulation at 300 K. The hydrostatic stress was calculated using the Virial theorem \cite{ZWHJKB2004} and was binned along the thickness direction. The surface stress was calculated as the difference in hydrostatic stress between surface and bulk. This calculated surface stress is plotted as a function of position along the thickness in Fig. \ref{stress profile} for both modified and unmodified SW potentials. It can be seen that the modified potential clearly shows a significant surface stress compared with the unmodified potential. Thermal conductivities were calculated and we found $27.56 \pm 0.36 W/K \cdot m$ for the modified potential as compared with $27.70 \pm 0.43 W/K \cdot m$ for the unmodified potential. Based upon the data, we do not expect that neglecting the surface stress by the SW potential significantly affects the thermal conductivity estimates. 

\section{Conclusions}

We have explored general scaling equations that explicitly express thermal conductivity of film and wire as functions of dimensions. Based upon these scaling equations, we have demonstrated methods that enable molecular dynamics simulations to be used to predict thermal conductivity of nanostructures at realistic length scales. We have performed extensive MD calculations of thermal conducting properties along the $[0001]$ direction of GaN films and wires. The following conclusions have been obtained:

\begin{itemize}
\item The linear relationships predicted from the scaling equations hold extremely well for MD data at a large parameter $d$. Reliable prediction of film thermal conductivity as a function of film thickness has been achieved using linear interpolation.
\item Thermal flux in nanostructures exhibit a clear difference between the surface and core whereas temperature is nearly constant across the cross-section, thereby verifying the core-shell assumption of the scaling theory and the near-zero heat flux perpendicular to the axis connecting hot and cold reservoirs.
\item Due to the limitation of computational cost, parameters deduced from direct MD simulations of wires may not sufficiently accurately predict the wire thermal conductivity at large wire radii. However, the parameters deduced from film simulations enable the derivation of a reliable expression of wire thermal conductivity as a function of wire radius.
\item The simulated dimension does not affect the thermal conductivity when the dimension is transverse to the heat flow and a periodic boundary condition is used in that direction. Hence, the periodic boundary conditions can be used to accurately extend the system dimension to infinity.
\item The surface stress due to the contraction of surface bonds does not sensitively affect the thermal conductivity. As a result, the SW potential should be a sufficiently accurate force field for surface thermal transfer problems.
\end{itemize}

\begin{acknowledgments}

Sandia is a multi-program laboratory operated by Sandia Corporation, a Lockheed Martin Company, for the United States Department of Energy National Nuclear Security Administration under contract DEAC04-94AL85000. This work was performed under a Laboratory Directed Research and Development (LDRD) project.

\end{acknowledgments}
 
\appendix
 


\begin{thebibliography}{54}
\expandafter\ifx\csname natexlab\endcsname\relax\def\natexlab#1{#1}\fi
\expandafter\ifx\csname bibnamefont\endcsname\relax
  \def\bibnamefont#1{#1}\fi
\expandafter\ifx\csname bibfnamefont\endcsname\relax
  \def\bibfnamefont#1{#1}\fi
\expandafter\ifx\csname citenamefont\endcsname\relax
  \def\citenamefont#1{#1}\fi
\expandafter\ifx\csname url\endcsname\relax
  \def\url#1{\texttt{#1}}\fi
\expandafter\ifx\csname urlprefix\endcsname\relax\def\urlprefix{URL }\fi
\providecommand{\bibinfo}[2]{#2}
\providecommand{\eprint}[2][]{\url{#2}}

\bibitem[{\citenamefont{Shakouri}(2006)}]{S2006}
\bibinfo{author}{\bibfnamefont{A.}~\bibnamefont{Shakouri}},
  \bibinfo{journal}{Proc. IEEE} \textbf{\bibinfo{volume}{94}},
  \bibinfo{pages}{1613} (\bibinfo{year}{2006}).

\bibitem[{\citenamefont{Balandin and Wang}(1998)}]{BW1998}
\bibinfo{author}{\bibfnamefont{A.}~\bibnamefont{Balandin}} \bibnamefont{and}
  \bibinfo{author}{\bibfnamefont{K.~L.} \bibnamefont{Wang}},
  \bibinfo{journal}{Phys. Rev. B} \textbf{\bibinfo{volume}{58}},
  \bibinfo{pages}{1544} (\bibinfo{year}{1998}).

\bibitem[{\citenamefont{Zou and Balandin}(2001)}]{ZB2001}
\bibinfo{author}{\bibfnamefont{J.}~\bibnamefont{Zou}} \bibnamefont{and}
  \bibinfo{author}{\bibfnamefont{A.}~\bibnamefont{Balandin}},
  \bibinfo{journal}{J. Appl. Phys.} \textbf{\bibinfo{volume}{89}},
  \bibinfo{pages}{2932} (\bibinfo{year}{2001}).

\bibitem[{\citenamefont{Goodson and Ju}(1999)}]{GJ1999}
\bibinfo{author}{\bibfnamefont{K.~E.} \bibnamefont{Goodson}} \bibnamefont{and}
  \bibinfo{author}{\bibfnamefont{Y.~S.} \bibnamefont{Ju}},
  \bibinfo{journal}{Ann. Rev. Mater. Sci.} \textbf{\bibinfo{volume}{29}},
  \bibinfo{pages}{261} (\bibinfo{year}{1999}).

\bibitem[{\citenamefont{Balandin}(2000)}]{B2000}
\bibinfo{author}{\bibfnamefont{A.}~\bibnamefont{Balandin}},
  \bibinfo{journal}{Phys. Low-Dimen. Struct.} \textbf{\bibinfo{volume}{1-2}},
  \bibinfo{pages}{1} (\bibinfo{year}{2000}).

\bibitem[{\citenamefont{Mahan et~al.}(1997)\citenamefont{Mahan, Sales, and
  Sharp}}]{MSS1997}
\bibinfo{author}{\bibfnamefont{G.}~\bibnamefont{Mahan}},
  \bibinfo{author}{\bibfnamefont{B.}~\bibnamefont{Sales}}, \bibnamefont{and}
  \bibinfo{author}{\bibfnamefont{J.}~\bibnamefont{Sharp}},
  \bibinfo{journal}{Phys. Today} \textbf{\bibinfo{volume}{50}},
  \bibinfo{pages}{42} (\bibinfo{year}{1997}).

\bibitem[{\citenamefont{Zhang}(2007)}]{Zhang2007}
\bibinfo{author}{\bibfnamefont{Z.~M.} \bibnamefont{Zhang}},
  \emph{\bibinfo{title}{Nano/microscale heat transfer}}
  (\bibinfo{publisher}{McGraw-Hill}, \bibinfo{address}{New York},
  \bibinfo{year}{2007}).

\bibitem[{\citenamefont{Lu et~al.}(2001)\citenamefont{Lu, Gu, and
  Chu}}]{LLGC2001}
\bibinfo{author}{\bibfnamefont{X.}~\bibnamefont{Lu}},
  \bibinfo{author}{\bibfnamefont{J.~H.} \bibnamefont{Gu}}, \bibnamefont{and}
  \bibinfo{author}{\bibfnamefont{J.~H.} \bibnamefont{Chu}},
  \bibinfo{journal}{Chin. Phys. Soc.} \textbf{\bibinfo{volume}{10}},
  \bibinfo{pages}{223} (\bibinfo{year}{2001}).

\bibitem[{\citenamefont{Lu et~al.}(2002)\citenamefont{Lu, Shen, and
  Chu}}]{LSC2002}
\bibinfo{author}{\bibfnamefont{X.}~\bibnamefont{Lu}},
  \bibinfo{author}{\bibfnamefont{W.~S.} \bibnamefont{Shen}}, \bibnamefont{and}
  \bibinfo{author}{\bibfnamefont{J.~H.} \bibnamefont{Chu}},
  \bibinfo{journal}{J. Appl. Phys.} \textbf{\bibinfo{volume}{91}},
  \bibinfo{pages}{1542} (\bibinfo{year}{2002}).

\bibitem[{\citenamefont{Walkauskas et~al.}(1999)\citenamefont{Walkauskas,
  Broido, Kempa, and Reinecke}}]{WBKR1999}
\bibinfo{author}{\bibfnamefont{S.~G.} \bibnamefont{Walkauskas}},
  \bibinfo{author}{\bibfnamefont{D.~A.} \bibnamefont{Broido}},
  \bibinfo{author}{\bibfnamefont{K.}~\bibnamefont{Kempa}}, \bibnamefont{and}
  \bibinfo{author}{\bibfnamefont{T.~L.} \bibnamefont{Reinecke}},
  \bibinfo{journal}{J. Appl. Phys.} \textbf{\bibinfo{volume}{85}},
  \bibinfo{pages}{2579} (\bibinfo{year}{1999}).

\bibitem[{\citenamefont{Mingo and Broido}(2004)}]{MB2004}
\bibinfo{author}{\bibfnamefont{N.}~\bibnamefont{Mingo}} \bibnamefont{and}
  \bibinfo{author}{\bibfnamefont{D.~A.} \bibnamefont{Broido}},
  \bibinfo{journal}{Phys. Rev. Lett.} \textbf{\bibinfo{volume}{93}},
  \bibinfo{pages}{246106} (\bibinfo{year}{2004}).

\bibitem[{\citenamefont{Volz and Chen}(1999)}]{VC1999}
\bibinfo{author}{\bibfnamefont{S.~G.} \bibnamefont{Volz}} \bibnamefont{and}
  \bibinfo{author}{\bibfnamefont{G.}~\bibnamefont{Chen}},
  \bibinfo{journal}{Appl. Phys. Lett.} \textbf{\bibinfo{volume}{75}},
  \bibinfo{pages}{2056} (\bibinfo{year}{1999}).

\bibitem[{\citenamefont{Kawamura et~al.}(2005)\citenamefont{Kawamura, Kangawa,
  and Kakimoto}}]{KKK2005}
\bibinfo{author}{\bibfnamefont{T.}~\bibnamefont{Kawamura}},
  \bibinfo{author}{\bibfnamefont{Y.}~\bibnamefont{Kangawa}}, \bibnamefont{and}
  \bibinfo{author}{\bibfnamefont{K.}~\bibnamefont{Kakimoto}},
  \bibinfo{journal}{J. Cryst. Growth} \textbf{\bibinfo{volume}{284}},
  \bibinfo{pages}{197} (\bibinfo{year}{2005}).

\bibitem[{\citenamefont{Che et~al.}(2000{\natexlab{a}})\citenamefont{Che,
  Cagin, Deng, and Goddard}}]{CCDG2000}
\bibinfo{author}{\bibfnamefont{J.~W.} \bibnamefont{Che}},
  \bibinfo{author}{\bibfnamefont{T.}~\bibnamefont{Cagin}},
  \bibinfo{author}{\bibfnamefont{W.~Q.} \bibnamefont{Deng}}, \bibnamefont{and}
  \bibinfo{author}{\bibfnamefont{W.~A.} \bibnamefont{Goddard}},
  \bibinfo{journal}{J. Chem. Phys.} \textbf{\bibinfo{volume}{113}},
  \bibinfo{pages}{6888} (\bibinfo{year}{2000}{\natexlab{a}}).

\bibitem[{\citenamefont{Che et~al.}(2000{\natexlab{b}})\citenamefont{Che,
  Cagin, and III}}]{CCG2000}
\bibinfo{author}{\bibfnamefont{J.~W.} \bibnamefont{Che}},
  \bibinfo{author}{\bibfnamefont{T.}~\bibnamefont{Cagin}}, \bibnamefont{and}
  \bibinfo{author}{\bibfnamefont{W.~A.~G.} \bibnamefont{III}},
  \bibinfo{journal}{Nanotechnology} \textbf{\bibinfo{volume}{11}},
  \bibinfo{pages}{65} (\bibinfo{year}{2000}{\natexlab{b}}).

\bibitem[{\citenamefont{Li et~al.}(1998)\citenamefont{Li, Porter, and
  Yip}}]{LPY1998}
\bibinfo{author}{\bibfnamefont{J.}~\bibnamefont{Li}},
  \bibinfo{author}{\bibfnamefont{L.}~\bibnamefont{Porter}}, \bibnamefont{and}
  \bibinfo{author}{\bibfnamefont{S.}~\bibnamefont{Yip}}, \bibinfo{journal}{J.
  Nucl. Mater.} \textbf{\bibinfo{volume}{255}}, \bibinfo{pages}{139}
  (\bibinfo{year}{1998}).

\bibitem[{\citenamefont{Volz and Chen}(2000)}]{VC2000}
\bibinfo{author}{\bibfnamefont{S.~G.} \bibnamefont{Volz}} \bibnamefont{and}
  \bibinfo{author}{\bibfnamefont{G.}~\bibnamefont{Chen}},
  \bibinfo{journal}{Phys. Rev. B} \textbf{\bibinfo{volume}{61}},
  \bibinfo{pages}{2651} (\bibinfo{year}{2000}).

\bibitem[{\citenamefont{Ladd et~al.}(1986)\citenamefont{Ladd, Moran, and
  Hoover}}]{LMH1986}
\bibinfo{author}{\bibfnamefont{A.~J.~C.} \bibnamefont{Ladd}},
  \bibinfo{author}{\bibfnamefont{B.}~\bibnamefont{Moran}}, \bibnamefont{and}
  \bibinfo{author}{\bibfnamefont{W.~G.} \bibnamefont{Hoover}},
  \bibinfo{journal}{Phys. Rev. B} \textbf{\bibinfo{volume}{34}},
  \bibinfo{pages}{5058} (\bibinfo{year}{1986}).

\bibitem[{\citenamefont{Vogelsang et~al.}(1987)\citenamefont{Vogelsang,
  Hoheisel, and Ciccotti}}]{VHC1987}
\bibinfo{author}{\bibfnamefont{R.}~\bibnamefont{Vogelsang}},
  \bibinfo{author}{\bibfnamefont{C.}~\bibnamefont{Hoheisel}}, \bibnamefont{and}
  \bibinfo{author}{\bibfnamefont{G.}~\bibnamefont{Ciccotti}},
  \bibinfo{journal}{J. Chem. Phys.} \textbf{\bibinfo{volume}{86}},
  \bibinfo{pages}{6371} (\bibinfo{year}{1987}).

\bibitem[{\citenamefont{Maiti et~al.}(1997)\citenamefont{Maiti, Mahan, and
  Pantelides}}]{MMP1997}
\bibinfo{author}{\bibfnamefont{A.}~\bibnamefont{Maiti}},
  \bibinfo{author}{\bibfnamefont{G.~D.} \bibnamefont{Mahan}}, \bibnamefont{and}
  \bibinfo{author}{\bibfnamefont{S.~T.} \bibnamefont{Pantelides}},
  \bibinfo{journal}{Solid State Commun.} \textbf{\bibinfo{volume}{102}},
  \bibinfo{pages}{517} (\bibinfo{year}{1997}).

\bibitem[{\citenamefont{Oligschleger and Schon}(1999)}]{OS1999}
\bibinfo{author}{\bibfnamefont{C.}~\bibnamefont{Oligschleger}}
  \bibnamefont{and} \bibinfo{author}{\bibfnamefont{J.~C.} \bibnamefont{Schon}},
  \bibinfo{journal}{Phys. Rev. B} \textbf{\bibinfo{volume}{59}},
  \bibinfo{pages}{4125} (\bibinfo{year}{1999}).

\bibitem[{\citenamefont{Michalski}(1992)}]{M1992}
\bibinfo{author}{\bibfnamefont{J.}~\bibnamefont{Michalski}},
  \bibinfo{journal}{Phys. Rev. B} \textbf{\bibinfo{volume}{45}},
  \bibinfo{pages}{7054} (\bibinfo{year}{1992}).

\bibitem[{\citenamefont{Poetzsch and Bottger}(1994)}]{PB1994}
\bibinfo{author}{\bibfnamefont{R.~H.~H.} \bibnamefont{Poetzsch}}
  \bibnamefont{and} \bibinfo{author}{\bibfnamefont{H.}~\bibnamefont{Bottger}},
  \bibinfo{journal}{Phys. Rev. B} \textbf{\bibinfo{volume}{50}},
  \bibinfo{pages}{15757} (\bibinfo{year}{1994}).

\bibitem[{\citenamefont{Baranyai}(1996)}]{B1996}
\bibinfo{author}{\bibfnamefont{A.}~\bibnamefont{Baranyai}},
  \bibinfo{journal}{Phys. Rev. E} \textbf{\bibinfo{volume}{54}},
  \bibinfo{pages}{6911} (\bibinfo{year}{1996}).

\bibitem[{\citenamefont{Schelling and Phillpot}(2001)}]{SP2001}
\bibinfo{author}{\bibfnamefont{P.~K.} \bibnamefont{Schelling}}
  \bibnamefont{and} \bibinfo{author}{\bibfnamefont{S.~R.}
  \bibnamefont{Phillpot}}, \bibinfo{journal}{J. Am. Ceram. Soc.}
  \textbf{\bibinfo{volume}{84}}, \bibinfo{pages}{2997} (\bibinfo{year}{2001}).

\bibitem[{\citenamefont{Jund and Jullien}(1999)}]{JJ1999}
\bibinfo{author}{\bibfnamefont{P.}~\bibnamefont{Jund}} \bibnamefont{and}
  \bibinfo{author}{\bibfnamefont{R.}~\bibnamefont{Jullien}},
  \bibinfo{journal}{Phys. Rev. B} \textbf{\bibinfo{volume}{59}},
  \bibinfo{pages}{13707} (\bibinfo{year}{1999}).

\bibitem[{\citenamefont{Schelling et~al.}(2002)\citenamefont{Schelling,
  Phillpot, and Keblinski}}]{SPK2002}
\bibinfo{author}{\bibfnamefont{P.~K.} \bibnamefont{Schelling}},
  \bibinfo{author}{\bibfnamefont{S.~R.} \bibnamefont{Phillpot}},
  \bibnamefont{and}
  \bibinfo{author}{\bibfnamefont{P.}~\bibnamefont{Keblinski}},
  \bibinfo{journal}{Phys. Rev. B} \textbf{\bibinfo{volume}{65}},
  \bibinfo{pages}{144306} (\bibinfo{year}{2002}).

\bibitem[{\citenamefont{Schelling et~al.}(2004)\citenamefont{Schelling,
  Phillpot, and Keblinski}}]{SPK2004}
\bibinfo{author}{\bibfnamefont{P.~K.} \bibnamefont{Schelling}},
  \bibinfo{author}{\bibfnamefont{S.~R.} \bibnamefont{Phillpot}},
  \bibnamefont{and}
  \bibinfo{author}{\bibfnamefont{P.}~\bibnamefont{Keblinski}},
  \bibinfo{journal}{J. Appl. Phys.} \textbf{\bibinfo{volume}{95}},
  \bibinfo{pages}{6082} (\bibinfo{year}{2004}).

\bibitem[{\citenamefont{Yoon et~al.}(2004)\citenamefont{Yoon, Car, Srolovitz,
  and Scandolo}}]{YCSS2004}
\bibinfo{author}{\bibfnamefont{Y.-G.} \bibnamefont{Yoon}},
  \bibinfo{author}{\bibfnamefont{R.}~\bibnamefont{Car}},
  \bibinfo{author}{\bibfnamefont{D.~J.} \bibnamefont{Srolovitz}},
  \bibnamefont{and} \bibinfo{author}{\bibfnamefont{S.}~\bibnamefont{Scandolo}},
  \bibinfo{journal}{Phys. Rev. B} \textbf{\bibinfo{volume}{70}},
  \bibinfo{pages}{012302} (\bibinfo{year}{2004}).

\bibitem[{\citenamefont{Wang et~al.}(2009)\citenamefont{Wang, Liang, Xu, and
  Ohara}}]{WLXO2009}
\bibinfo{author}{\bibfnamefont{S.-C.} \bibnamefont{Wang}},
  \bibinfo{author}{\bibfnamefont{X.-G.} \bibnamefont{Liang}},
  \bibinfo{author}{\bibfnamefont{X.-H.} \bibnamefont{Xu}}, \bibnamefont{and}
  \bibinfo{author}{\bibfnamefont{T.}~\bibnamefont{Ohara}}, \bibinfo{journal}{J.
  Appl. Phys.} \textbf{\bibinfo{volume}{105}}, \bibinfo{pages}{014316}
  (\bibinfo{year}{2009}).

\bibitem[{\citenamefont{Zhou et~al.}(2009)\citenamefont{Zhou, Aubry, Jones,
  Greenstein, and Schelling}}]{ZAJGS2009}
\bibinfo{author}{\bibfnamefont{X.~W.} \bibnamefont{Zhou}},
  \bibinfo{author}{\bibfnamefont{S.}~\bibnamefont{Aubry}},
  \bibinfo{author}{\bibfnamefont{R.~E.} \bibnamefont{Jones}},
  \bibinfo{author}{\bibfnamefont{A.}~\bibnamefont{Greenstein}},
  \bibnamefont{and} \bibinfo{author}{\bibfnamefont{P.~K.}
  \bibnamefont{Schelling}}, \bibinfo{journal}{Phys. Rev. B}
  \textbf{\bibinfo{volume}{79}}, \bibinfo{pages}{115201}
  (\bibinfo{year}{2009}).

\bibitem[{\citenamefont{Simpkins et~al.}(2007)\citenamefont{Simpkins, Pehrsson,
  Taheri, and Stroud}}]{SPTS2007}
\bibinfo{author}{\bibfnamefont{B.~S.} \bibnamefont{Simpkins}},
  \bibinfo{author}{\bibfnamefont{P.~E.} \bibnamefont{Pehrsson}},
  \bibinfo{author}{\bibfnamefont{M.~L.} \bibnamefont{Taheri}},
  \bibnamefont{and} \bibinfo{author}{\bibfnamefont{R.~M.}
  \bibnamefont{Stroud}}, \bibinfo{journal}{J. Appl. Phys.}
  \textbf{\bibinfo{volume}{101}}, \bibinfo{pages}{094305}
  (\bibinfo{year}{2007}).

\bibitem[{\citenamefont{Landry et~al.}(2008)\citenamefont{Landry, Hussein, and
  McGaughey}}]{LHM2008}
\bibinfo{author}{\bibfnamefont{E.~S.} \bibnamefont{Landry}},
  \bibinfo{author}{\bibfnamefont{M.~I.} \bibnamefont{Hussein}},
  \bibnamefont{and} \bibinfo{author}{\bibfnamefont{A.~J.~H.}
  \bibnamefont{McGaughey}}, \bibinfo{journal}{Phys. Rev. B}
  \textbf{\bibinfo{volume}{77}}, \bibinfo{pages}{184302}
  (\bibinfo{year}{2008}).

\bibitem[{\citenamefont{Heino}(2007)}]{H2007}
\bibinfo{author}{\bibfnamefont{P.}~\bibnamefont{Heino}}, \bibinfo{journal}{J.
  Comput. Theor. Nanosci.} \textbf{\bibinfo{volume}{4}}, \bibinfo{pages}{896}
  (\bibinfo{year}{2007}).

\bibitem[{\citenamefont{Ponomareva et~al.}(2007)\citenamefont{Ponomareva,
  Srivastava, and Menon}}]{PSM2007}
\bibinfo{author}{\bibfnamefont{I.}~\bibnamefont{Ponomareva}},
  \bibinfo{author}{\bibfnamefont{D.}~\bibnamefont{Srivastava}},
  \bibnamefont{and} \bibinfo{author}{\bibfnamefont{M.}~\bibnamefont{Menon}},
  \bibinfo{journal}{Nano Lett.} \textbf{\bibinfo{volume}{7}},
  \bibinfo{pages}{1155} (\bibinfo{year}{2007}).

\bibitem[{\citenamefont{Zhou et~al.}(2010)\citenamefont{Zhou, Jones, and
  Aubry}}]{ZJA2009}
\bibinfo{author}{\bibfnamefont{X.~W.} \bibnamefont{Zhou}},
  \bibinfo{author}{\bibfnamefont{R.~E.} \bibnamefont{Jones}}, \bibnamefont{and}
  \bibinfo{author}{\bibfnamefont{S.}~\bibnamefont{Aubry}},
  \bibinfo{journal}{Phys. Rev. B} \textbf{\bibinfo{volume}{81}},
  \bibinfo{pages}{073304} (\bibinfo{year}{2010}).

\bibitem[{\citenamefont{Johnson et~al.}(2002)\citenamefont{Johnson, Choi,
  Knutsen, Schaller, Yang, and Saykally}}]{JCKSYS2002}
\bibinfo{author}{\bibfnamefont{J.~C.} \bibnamefont{Johnson}},
  \bibinfo{author}{\bibfnamefont{H.~J.} \bibnamefont{Choi}},
  \bibinfo{author}{\bibfnamefont{K.~P.} \bibnamefont{Knutsen}},
  \bibinfo{author}{\bibfnamefont{R.~D.} \bibnamefont{Schaller}},
  \bibinfo{author}{\bibfnamefont{P.~D.} \bibnamefont{Yang}}, \bibnamefont{and}
  \bibinfo{author}{\bibfnamefont{R.~J.} \bibnamefont{Saykally}},
  \bibinfo{journal}{Nature Mater.} \textbf{\bibinfo{volume}{1}},
  \bibinfo{pages}{106} (\bibinfo{year}{2002}).

\bibitem[{\citenamefont{Zhong et~al.}(2003)\citenamefont{Zhong, Qian, Wang, and
  Lieber}}]{ZQWL2003}
\bibinfo{author}{\bibfnamefont{Z.~H.} \bibnamefont{Zhong}},
  \bibinfo{author}{\bibfnamefont{F.}~\bibnamefont{Qian}},
  \bibinfo{author}{\bibfnamefont{D.~L.} \bibnamefont{Wang}}, \bibnamefont{and}
  \bibinfo{author}{\bibfnamefont{C.~M.} \bibnamefont{Lieber}},
  \bibinfo{journal}{Nano Lett.} \textbf{\bibinfo{volume}{3}},
  \bibinfo{pages}{343} (\bibinfo{year}{2003}).

\bibitem[{\citenamefont{Kim et~al.}(2004)\citenamefont{Kim, Cho, Lee, Kim, Ryu,
  Kim, Kang, and Chung}}]{KCLKRKKC2004}
\bibinfo{author}{\bibfnamefont{H.~M.} \bibnamefont{Kim}},
  \bibinfo{author}{\bibfnamefont{Y.~H.} \bibnamefont{Cho}},
  \bibinfo{author}{\bibfnamefont{H.}~\bibnamefont{Lee}},
  \bibinfo{author}{\bibfnamefont{S.~I.} \bibnamefont{Kim}},
  \bibinfo{author}{\bibfnamefont{S.~R.} \bibnamefont{Ryu}},
  \bibinfo{author}{\bibfnamefont{D.~Y.} \bibnamefont{Kim}},
  \bibinfo{author}{\bibfnamefont{T.~W.} \bibnamefont{Kang}}, \bibnamefont{and}
  \bibinfo{author}{\bibfnamefont{K.~S.} \bibnamefont{Chung}},
  \bibinfo{journal}{Nano Lett.} \textbf{\bibinfo{volume}{4}},
  \bibinfo{pages}{1059} (\bibinfo{year}{2004}).

\bibitem[{\citenamefont{Qian et~al.}(2004)\citenamefont{Qian, Li, Gradecak,
  Wang, Barrelet, and Lieber}}]{QLGWBL2004}
\bibinfo{author}{\bibfnamefont{F.}~\bibnamefont{Qian}},
  \bibinfo{author}{\bibfnamefont{Y.}~\bibnamefont{Li}},
  \bibinfo{author}{\bibfnamefont{S.}~\bibnamefont{Gradecak}},
  \bibinfo{author}{\bibfnamefont{D.~L.} \bibnamefont{Wang}},
  \bibinfo{author}{\bibfnamefont{C.~J.} \bibnamefont{Barrelet}},
  \bibnamefont{and} \bibinfo{author}{\bibfnamefont{C.~M.}
  \bibnamefont{Lieber}}, \bibinfo{journal}{Nano Lett.}
  \textbf{\bibinfo{volume}{4}}, \bibinfo{pages}{1975} (\bibinfo{year}{2004}).

\bibitem[{\citenamefont{Huang et~al.}(2002)\citenamefont{Huang, Duan, Cui, and
  Lieber}}]{HDCL2002}
\bibinfo{author}{\bibfnamefont{Y.}~\bibnamefont{Huang}},
  \bibinfo{author}{\bibfnamefont{X.~F.} \bibnamefont{Duan}},
  \bibinfo{author}{\bibfnamefont{Y.}~\bibnamefont{Cui}}, \bibnamefont{and}
  \bibinfo{author}{\bibfnamefont{C.~M.} \bibnamefont{Lieber}},
  \bibinfo{journal}{Nano Lett.} \textbf{\bibinfo{volume}{2}},
  \bibinfo{pages}{101} (\bibinfo{year}{2002}).

\bibitem[{\citenamefont{Choi et~al.}(2003)\citenamefont{Choi, Johnson, He, Lee,
  Kim, Pauzauskie, Goldberger, Saykally, and Yang}}]{CJHLKPGSY2003}
\bibinfo{author}{\bibfnamefont{H.~J.} \bibnamefont{Choi}},
  \bibinfo{author}{\bibfnamefont{J.~C.} \bibnamefont{Johnson}},
  \bibinfo{author}{\bibfnamefont{R.~R.} \bibnamefont{He}},
  \bibinfo{author}{\bibfnamefont{S.~K.} \bibnamefont{Lee}},
  \bibinfo{author}{\bibfnamefont{F.}~\bibnamefont{Kim}},
  \bibinfo{author}{\bibfnamefont{P.}~\bibnamefont{Pauzauskie}},
  \bibinfo{author}{\bibfnamefont{J.}~\bibnamefont{Goldberger}},
  \bibinfo{author}{\bibfnamefont{R.~J.} \bibnamefont{Saykally}},
  \bibnamefont{and} \bibinfo{author}{\bibfnamefont{P.~D.} \bibnamefont{Yang}},
  \bibinfo{journal}{J. Phys. Chem. B} \textbf{\bibinfo{volume}{107}},
  \bibinfo{pages}{8721} (\bibinfo{year}{2003}).

\bibitem[{\citenamefont{Danilchenko et~al.}(2006)\citenamefont{Danilchenko,
  Paszkiewicz, Wolski, Jezowski, and Plackowski}}]{DPWJP2006}
\bibinfo{author}{\bibfnamefont{B.~A.} \bibnamefont{Danilchenko}},
  \bibinfo{author}{\bibfnamefont{T.}~\bibnamefont{Paszkiewicz}},
  \bibinfo{author}{\bibfnamefont{S.}~\bibnamefont{Wolski}},
  \bibinfo{author}{\bibfnamefont{A.}~\bibnamefont{Jezowski}}, \bibnamefont{and}
  \bibinfo{author}{\bibfnamefont{T.}~\bibnamefont{Plackowski}},
  \bibinfo{journal}{Appl. Phys. Lett.} \textbf{\bibinfo{volume}{89}},
  \bibinfo{pages}{061901} (\bibinfo{year}{2006}).

\bibitem[{\citenamefont{Bere and Serra}(2002)}]{BS2002}
\bibinfo{author}{\bibfnamefont{A.}~\bibnamefont{Bere}} \bibnamefont{and}
  \bibinfo{author}{\bibfnamefont{A.}~\bibnamefont{Serra}},
  \bibinfo{journal}{Phys. Rev. B} \textbf{\bibinfo{volume}{65}},
  \bibinfo{pages}{205323} (\bibinfo{year}{2002}).

\bibitem[{\citenamefont{Bere and Serra}(2006)}]{BS2006}
\bibinfo{author}{\bibfnamefont{A.}~\bibnamefont{Bere}} \bibnamefont{and}
  \bibinfo{author}{\bibfnamefont{A.}~\bibnamefont{Serra}},
  \bibinfo{journal}{Phil. Mag.} \textbf{\bibinfo{volume}{86}},
  \bibinfo{pages}{2159} (\bibinfo{year}{2006}).

\bibitem[{\citenamefont{Serrano et~al.}(2000)\citenamefont{Serrano, Rubio,
  Hern\'{a}ndez, Mu$\tilde{n}$oz, and Mujica}}]{SRHMM2000}
\bibinfo{author}{\bibfnamefont{J.}~\bibnamefont{Serrano}},
  \bibinfo{author}{\bibfnamefont{A.}~\bibnamefont{Rubio}},
  \bibinfo{author}{\bibfnamefont{E.}~\bibnamefont{Hern\'{a}ndez}},
  \bibinfo{author}{\bibfnamefont{A.}~\bibnamefont{Mu$\tilde{n}$oz}},
  \bibnamefont{and} \bibinfo{author}{\bibfnamefont{A.}~\bibnamefont{Mujica}},
  \bibinfo{journal}{Phys. Rev. B} \textbf{\bibinfo{volume}{62}},
  \bibinfo{pages}{16612} (\bibinfo{year}{2000}).

\bibitem[{\citenamefont{Bertelli et~al.}(2009)\citenamefont{Bertelli, Loptien,
  Wenderoth, Rizzi, Ulbrich, Righi, Ferretti, Martin-Samos, Bertoni, and
  Catellani}}]{BLWRURFMBC2009}
\bibinfo{author}{\bibfnamefont{M.}~\bibnamefont{Bertelli}},
  \bibinfo{author}{\bibfnamefont{P.}~\bibnamefont{Loptien}},
  \bibinfo{author}{\bibfnamefont{M.}~\bibnamefont{Wenderoth}},
  \bibinfo{author}{\bibfnamefont{A.}~\bibnamefont{Rizzi}},
  \bibinfo{author}{\bibfnamefont{R.~G.} \bibnamefont{Ulbrich}},
  \bibinfo{author}{\bibfnamefont{M.~C.} \bibnamefont{Righi}},
  \bibinfo{author}{\bibfnamefont{A.}~\bibnamefont{Ferretti}},
  \bibinfo{author}{\bibfnamefont{L.}~\bibnamefont{Martin-Samos}},
  \bibinfo{author}{\bibfnamefont{C.~M.} \bibnamefont{Bertoni}},
  \bibnamefont{and}
  \bibinfo{author}{\bibfnamefont{A.}~\bibnamefont{Catellani}},
  \bibinfo{journal}{Phys. Rev. B} \textbf{\bibinfo{volume}{80}},
  \bibinfo{pages}{115324} (\bibinfo{year}{2009}).

\bibitem[{\citenamefont{Ikeshoji and Hafskjold}(1994)}]{IH1994}
\bibinfo{author}{\bibfnamefont{T.}~\bibnamefont{Ikeshoji}} \bibnamefont{and}
  \bibinfo{author}{\bibfnamefont{B.}~\bibnamefont{Hafskjold}},
  \bibinfo{journal}{Mol. Phys.} \textbf{\bibinfo{volume}{81}},
  \bibinfo{pages}{251} (\bibinfo{year}{1994}).

\bibitem[{\citenamefont{III et~al.}(2008)\citenamefont{III, Zimmerman, and
  Seel}}]{WZS2008}
\bibinfo{author}{\bibfnamefont{E.~B.~W.} \bibnamefont{III}},
  \bibinfo{author}{\bibfnamefont{J.~A.} \bibnamefont{Zimmerman}},
  \bibnamefont{and} \bibinfo{author}{\bibfnamefont{S.~C.} \bibnamefont{Seel}},
  \bibinfo{journal}{Math. Mech. Sol.} \textbf{\bibinfo{volume}{13}},
  \bibinfo{pages}{221} (\bibinfo{year}{2008}).

\bibitem[{\citenamefont{Slack}(1973)}]{S1973}
\bibinfo{author}{\bibfnamefont{G.~A.} \bibnamefont{Slack}},
  \bibinfo{journal}{J. Phys. Chem. Sol.} \textbf{\bibinfo{volume}{34}},
  \bibinfo{pages}{321} (\bibinfo{year}{1973}).

\bibitem[{\citenamefont{Marmalyuk et~al.}(1998)\citenamefont{Marmalyuk,
  Akchurin, and Gorbylev}}]{MAG1998}
\bibinfo{author}{\bibfnamefont{A.~A.} \bibnamefont{Marmalyuk}},
  \bibinfo{author}{\bibfnamefont{R.~K.} \bibnamefont{Akchurin}},
  \bibnamefont{and} \bibinfo{author}{\bibfnamefont{V.~A.}
  \bibnamefont{Gorbylev}}, \bibinfo{journal}{High Temperature}
  \textbf{\bibinfo{volume}{36}}, \bibinfo{pages}{817} (\bibinfo{year}{1998}).

\bibitem[{\citenamefont{Balamane et~al.}(1992)\citenamefont{Balamane,
  Halicioglu, and Tiller}}]{BHT1992}
\bibinfo{author}{\bibfnamefont{H.}~\bibnamefont{Balamane}},
  \bibinfo{author}{\bibfnamefont{T.}~\bibnamefont{Halicioglu}},
  \bibnamefont{and} \bibinfo{author}{\bibfnamefont{W.~A.}
  \bibnamefont{Tiller}}, \bibinfo{journal}{Phys. Rev. B}
  \textbf{\bibinfo{volume}{46}}, \bibinfo{pages}{2250} (\bibinfo{year}{1992}).

\bibitem[{\citenamefont{Nord et~al.}(2003)\citenamefont{Nord, Albe, Erhart, and
  Nordlund}}]{NAEN2003}
\bibinfo{author}{\bibfnamefont{J.}~\bibnamefont{Nord}},
  \bibinfo{author}{\bibfnamefont{K.}~\bibnamefont{Albe}},
  \bibinfo{author}{\bibfnamefont{P.}~\bibnamefont{Erhart}}, \bibnamefont{and}
  \bibinfo{author}{\bibfnamefont{K.}~\bibnamefont{Nordlund}},
  \bibinfo{journal}{J. Phys.} \textbf{\bibinfo{volume}{15}},
  \bibinfo{pages}{5649} (\bibinfo{year}{2003}).

\bibitem[{\citenamefont{Zimmerman et~al.}(2004)\citenamefont{Zimmerman, III,
  Hoyt, Jones, Klein, and Bammann}}]{ZWHJKB2004}
\bibinfo{author}{\bibfnamefont{J.}~\bibnamefont{Zimmerman}},
  \bibinfo{author}{\bibfnamefont{E.~W.} \bibnamefont{III}},
  \bibinfo{author}{\bibfnamefont{J.}~\bibnamefont{Hoyt}},
  \bibinfo{author}{\bibfnamefont{R.}~\bibnamefont{Jones}},
  \bibinfo{author}{\bibfnamefont{P.}~\bibnamefont{Klein}}, \bibnamefont{and}
  \bibinfo{author}{\bibfnamefont{D.}~\bibnamefont{Bammann}},
  \bibinfo{journal}{Modelling Simul. Mater. Sci. Eng.}
  \textbf{\bibinfo{volume}{12}}, \bibinfo{pages}{S319} (\bibinfo{year}{2004}).

\end{thebibliography}
\end{document}